\documentclass[pra,twocolumn,unsortedaddress,showpacs]{revtex4}
\usepackage{graphicx}
\usepackage{dcolumn}
\begin{document}

\title{Number--conserving model for boson pairing}

\author{S. Fantoni}
\affiliation{
International School for Advanced Studies, SISSA, 
I-34014 Trieste, Italy}
\affiliation{
International Centre for Theoretical Physics, ICTP, 
I-34014 Trieste, Italy}
\author{T. M. Nguyen}
\affiliation{
International Centre for Theoretical Physics, ICTP, 
I-34014 Trieste, Italy}
\author{A. Sarsa}
\affiliation{
International School for Advanced Studies, SISSA, 
I-34014 Trieste, Italy}
\author{S. R. Shenoy}
\affiliation{
International Centre for Theoretical Physics, ICTP, 
I-34014 Trieste, Italy}

\date{\today}

\begin{abstract}

An independent pair ansatz is developed for the many body 
wavefunction of dilute Bose systems. The pair correlation is 
optimized by minimizing the expectation value of the full hamiltonian 
(rather than the truncated  Bogoliubov one) providing a rigorous 
energy upper bound. In contrast with the Jastrow model, hypernetted 
chain theory provides closed-form exactly solvable equations for 
the optimized pair correlation. The model involves both condensate and 
coherent pairing with number conservation and kinetic energy sum rules 
satisfied exactly and the compressibility sum rule obeyed at low density.
We compute, for  bulk boson matter at a given density and zero
temperature, (i) the two--body distribution function, (ii) the 
energy per particle,  (iii) the sound velocity, (iv) the chemical 
potential, (v) the momentum distribution and its condensate fraction 
and (vi) the pairing function, which quantifies the ODLRO resulting 
from the structural properties of the two--particle density matrix. 
The connections with the low--density expansion and Bogoliubov theory 
are analyzed at different density values, including the density and 
scattering length regime of interest of trapped-atoms Bose--Einstein 
condensates. Comparison with the available Diffusion Monte Carlo results 
is also made.


\end{abstract}
 
\maketitle

\section{Introduction}

Models for interacting bosons, a focus of interest for decades,
have gained renewed importance due to the intense current activity in
Bose-Einstein condensates \cite{BEC,dalfovo99,ketterle99,leggett01}. 

The Bogoliubov model for weakly interacting 
bosons \cite{bogoliubov47,huang,feenberg} has
long been the starting point for more detailed 
theories \cite{girardeau59,valatin58,evans69,shenoy77,bohm67}.
The ground state is approximated by an ideal
Bose gas, and number conservation is violated by anomalous operator
averages of the zero momentum condensate operator.
Scattering out of (and depleting) the zero-momentum ground state  
produces opposite-momentum pairs that are treated as excitations 
with a linear spectrum \cite{bogoliubov47}.

Mean field pairing theories of superfluidity, by analogy with
superconductivity, considered BCS-like anomalous  pair--operator 
averages \cite{evans69,shenoy77}, 
preventing an undesirable BCS gap in the excitation
spectrum by a condition on the chemical potential \cite{pines59}.  
When neutron scattering estimated that in He II, the condensate 
fraction was less than 10\% \cite{henshaw61,nozieres99} 
while the superfluid fraction is 100\%, such
models were given added impetus, with pairing as a part (or whole) 
of the  ground state. More detailed many--body ground state 
calculations based on Jastrow wavefunctions go beyond the simple 
mean-field pairing, but to handle hard core potentials, must 
perforce have not only pairing, but also triplet, quadruplet and 
higher correlations. Thus many--body schemes like the hypernetted 
chain (HNC) approximation become open-ended hierarchies of coupled 
equations \cite{feenberg}.

On the other hand, the discovery of Bose-Einstein condensate 
(BEC) with a finite number of atoms prompted investigation of 
number-conserving reformulations of the Bogoliubov model, without
invoking anomalous averages, in modified Bogoliubov excitation 
operators \cite{leggett01,gardiner97}. Although the BEC depletion 
fraction is estimated to be small, it is nonzero, and issues remain, 
of a generalized understanding of the 100\% superfluid 
condensate \cite{leggett01}.

Motivated by low carrier-density superconductivity in high $T_c$ 
materials, negative-$U$ fermion pairing models have shown
\cite{legget69,randeria95} that \textit{real space} pairing 
(as in pre-BCS models \cite{valatin58,blatt64}) occurs at low densities, 
and goes over to momentum space BCS pairing at high densities.

In this paper, we consider a non-Jastrow ground state of 
\textit{number-conserving pairing} that ignores higher correlations, but
includes \textit{real space} pairing correlations \textit{exactly}. What 
we lose in a description of normal-state correlations and densities 
close to crystallization, we gain in a focus on the correlations 
responsible for superfluidity at lower densities. The emergent picture 
is of superfluidity from \textit{condensate plus real space pairs}, 
with the relative fraction dependent on density and interaction, 
but the superfluid fraction always unity at $T=0$. Within a pair-only 
model, the hypernetted chain scheme for  the ground state 
\textit{is now a closed form exactly soluble set of equations}, 
rather than being a correlation hierarchy,  to determine 
the two--body distribution function and the density matrices. 
Numerical solution of the 
Euler equation, resulting from the Ritz variational principle, 
yields equation of states for any given soft--core potential, 
and a momentum distribution with the correct long wavelength 
behavior, having the proper Gavoret--Nozi\'eres 
singularity \cite{nozieres64,stringari95}.

Correlation functions show off diagonal long-range order (ODLRO), 
factoring into pairing functions that
depend on the pair correlation defining the independent-pair ground state.
Number conservation and other important sum rules are
satisfied \cite{feenberg,nozieres99,stringari95}. 
The Bogoliubov--type results are recovered not at low densities,
but in the limit of large $\rho a ^3$, where $a$ is the 
inter-particle scattering length.

The interaction-induced presence of nonzero-momentum particles yields an 
occupation  number dependent wavevector that has 
some analogies with the ideal Fermi gas: one can define a nonzero 
\textit{Bose wavevector} $k_b$, an  average \textit{Bose energy} $E_b$,  
and a corresponding \textit{Bose temperature} $T_b$, 
that all vanish as a function of scaled
density/interaction, i.e. in the ideal Bose gas limit. 

More in detail, in this paper 
we use variational theory and Hyper Netted Chain 
methods \cite{feenberg,fantoni74,fantoni98}
to add pair correlations to the mean field
wave function, which is at the basis of the Gross--Pitaevsky
approach.  We consider the following ground state trial function 
which includes independent pair correlations (IPC) only

\begin{eqnarray}
\Psi_{IPC}(\vec{r}_1,\ldots ,\vec{r}_N) 
&=& 1 + \sum_{i < j} h(r_{ij}) \nonumber \\
 &+& \frac{1}{2!} \sum_{(i < j)\neq (l < m)}  h(r_{ij})  
h(r_{lm}) + \ldots,
\label{ipc}
 \end{eqnarray}

where the indices of the correlation factors $h(r_{ij})$, appearing
in the above summmed products, at all orders, never overlap. 
Writing $h(r)$ in Fourier space, one can easily verify that with the IPC
trial function of Eq.~(\ref{ipc}) the zero-momentum
condensate fraction interconverts with \textit{depletion} pairs with 
zero total momentum only. Note that $h$ is real i.e. condensation phase 
is locked to coherent pairing phases. Thus the $\Psi_{IPC}$ is characterized
by a condensate and a coherent depletion.

Although this condensate plus pair (only) ansatz is in the broad spirit of
the Bogoliubov model there are three crucial differences: i) Number
conservation is maintained, with no anomalous averages; ii) The 
condensate and pairs are coherent and both are part of the ground state
rather than the latter being excitations;
iii) The expectation valule of the \textit{total} hamiltonian is 
evaluated, without truncation, so the energy is a rigorous upper bound, and
the correlation $h(r)$ obtained by minimization goes beyond the
Bogoliubov approximation. Excitations are not considered here, but will
be related to density variations of the interaction-induced state of
coherent condensate-pair interconversion.

The IPC model constitutes the underlying wave function of Coupled Cluster
theory, in the SUB(2) approximation, which has been succesfully used in
a number of studies of strongly correlated 
systems \cite{navarro02,navarro02A}. In comparison with the Jastrow 
ansatz, $\Psi_J(R)=\prod_{i\le j}(1+h(r_{ij}))$, the correlation products
$h(r_{ij}) h(r_{lm})\ldots$ in Eq.~(\ref{ipc}) with $(ij) \neq (lm)$
constitute only a subset of those
produced by $\Psi_J(R)$, and, as a result, $\Psi_{IPC}(R)$ cannot be
put in a product form like $\Psi_J(R)$. Nonetheless, $\Psi_{IPC}(R)$,
still satisfies the separability condition, namely for any $
\{R\}=\{R_1,R_2\}$  
$\Psi_{IPC}(R_1,R_2)\rightarrow \Psi_{IPC}(R_1)\Psi_{IPC}(R_2)$, in
the limit of the subset of particles $\{R_1\}$ being macroscopically
far from $\{R_2\}$.

The trial function $\Psi_{IPC}(R)$ of Eq.~(\ref{ipc}), not being
of the product form,  
cannot deal directly with two--body potentials
which have strong singularities at short distances, like the
hard sphere potentials or those of the Lennard--Jones type. 
However, the detailed shape of the two--body 
potential may not be relevant for low-density systems, and suitable 
two--body  pseudopotentials \cite{aldrich76} can be found to study in 
a quantitative way the dynamical and superfluid properties of Bose systems.

We compute the two--body distribution function and one-- and two--particle 
density matrices of a bulk boson system by using a new HNC cluster 
technique, based on the Renormalized Fermi Hyper Netted Chain (RFHNC)
theory \cite{fantoni98}. This new cluster technique is made necessary 
because of the particular form of $\Psi_{IPC}(R)$, which
requires the summation of  
reducible and unlinked diagrams, contrary to the case of 
ordinary Jastrow theory. This novel method is based on treating 
the boson system as a collection of \textit{Fermi particles},
having the same mass and interaction of the
true bosonic ones, and a flavour degeneracy which equals 
the number of particles, so that they can all occupy the lowest state 
$\vec{k}=0$.
The above method relies on the property of the FHNC
cluster expansion to be exact at any order in $1/N$ 
\cite{fantoni74,fantoni98}. This could later permit the study of 
finite-atom trapped condensates.

The plan of this paper is as follows.
In Section II, we present the diagrammatics of the hypernetted chain
approximation, and show how the independent-pair model for the ground
state allows for a closed set of equations. In Section III we relate the
correlation functions (reduced or traced density matrix elements)
to the pair correlation $h(r)$ entering the ground state.
In Section IV we derive the Euler equations for the \textit{optimal} 
$h(r)$ and the corresponding momentum distribution $n(q)$. 
Results for different soft--core potentials, 
and plots of various physical quantities are given in Section V. 
Finally, Section VI is a summary and conclusions. 

\section{HNC theory for the independent pair model}

We present in this Section the calculation of the two--body
distribution function of a bulk boson system described by 
the IPC trial function of Eq.~(\ref{ipc}). 

The pair distribution function $g_2(1,2)$ 
is one of the fundamental quantities in the study of interacting
many--body systems.
Both the structure function $S(k)$ and the expectation value
of the hamiltonian $E[h]$ are given in terms of it. 
Therefore, its calculation is a necessary step for any quantitative
analysis of the static and dynamical property of an interacting quantum
many--body system. It is defined as

\begin{eqnarray}
g_2(1,2) &=& g(r_{12}) \nonumber \\
&=& \frac{N(N-1)}{\rho^2} \frac{\int d\vec{r}_3d\vec{r}_4\ldots
d\vec{r}_N |\Psi_{IPC}(R)|^2} {\int dR |\Psi_{IPC}(R)|^2} \ , 
\label{gdef}
  \end{eqnarray}

where the first equality is due to the translational invariance of the
system. The static structure function $S(k)$ is strictly related to
the pair function and given by \cite{feenberg}

\begin{eqnarray}
S(k) = 1+ \rho\int d\vec{r} e^{\imath \vec{k}\cdot \vec{r}} (g(r)-1) \ .
\label{stf}
\end{eqnarray}

The two--body distribution function $g(r)$ satisfies the following
normalization sum rule \cite{feenberg}

\begin{eqnarray}
S(0) = 1+\rho\int d\vec{r} (g(r)-1) = 0 \ . 
\label{grule}
  \end{eqnarray}
\subsection{The new HNC scheme for the IPC model.}

A new many--body method, based on the properties of the FHNC cluster
expansion \cite{fantoni74} and on the RFHNC summation 
technique \cite{fantoni98}, is developed and used in the calculation of
$g(r_{12})$. We show below, as done in Ref. \cite{fantoni98} for the
Jastrow model, that RFHNC applied to N-flavour fermions (which are
equivalent to a system of N bosons) produces the exact counting 
coefficients of the cluster diagrams in the independent pair ansatz.

\begin{figure}
\includegraphics{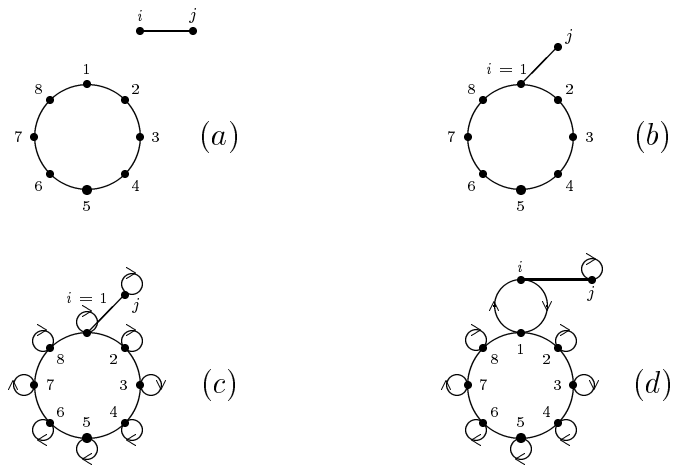}
\caption{\label{f1}
Diagrams showing the cancellation process of the HNC and the FHNC cluster
expansions as disussed in the text. All the four diagrams
give the same cluster integral $I$, and are constituted by the two
substructures $\Gamma(1,\ldots , s)$, with $s=8$, and $\gamma(i,j)$.
Diagrams (a) and (b) are bosonic. Diagrams (c) and (d) are fermionic, and
their oriented lines denote the exchange function $l(k_Fr)$.
}
\end{figure}

The standard HNC cluster decomposition is based on (i) a 
series expansion of the r.h.s. of Eq.~(\ref{gdef}) around $h(r)=0$ and (ii)
a diagrammatic representation of the various terms 
\footnote{For the Jastrow model the HNC cluster
decomposition coincides with the Mayer expansion.}.
The independent pair condition in $\Psi_{IPC}(R)$ implies that each 
vertex of any cluster diagram  
has at most two correlation lines reaching it, one
coming from  the ket $|\Psi_{IPC} \rangle$ ({K}--correlation) 
and the other coming from
the bra $\langle \Psi_{IPC}|$ ({B}--correlation).
This property, on the one hand, kills all the \textit{Jastrow} 
elementary diagrams, but, on the other, leads to a highly non trivial
cluster expansion with extra vertex corrections that can be summed. 
A new pairing-focussed HNC treatment, 
different from the standard one developed for 
Jastrow theory is needed to sum up the resulting cluster series, and
(unlike Jastrow) it can be written in closed form.

A first important subdivision of the cluster diagrams
is between \textit{linked} and \textit{unlinked} diagrams, where the
unlinked diagrams are those with two or more non overlapping parts. 
The linked diagrams are further subdivided in \textit{irreducible} and
\textit{reducible} diagrams. Reducible (or separable) diagrams correspond
to cluster integrals which can be factorized in two or more parts.  

In standard HNC, all the unlinked diagrams cancel each other, 
as schematically shown by the following simple example.
The upper part of Fig. \ref{f1} displays two cluster diagrams, 
(1a) and (1b) from different terms of the Mayer expansion.
Both are constituted with two linked 
diagrammatical structures, $\Gamma(1,\ldots,s)$ and
$\gamma(i,j)\equiv h(r_{ij})$, where $i$ or $j$ may or may not 
coincide with one of the labels of $\Gamma$. The structure 
$\Gamma(1,\ldots,s)$ has $s$ vertices, two of which correspond to
the two \textit{external} labels $1$ and $2$ of $g(r_{12})$; 
$\gamma(i,j)$ has two vertices only.  
Diagram (1a) is unlinked, whereas diagram (1b)
is linked but reducible. Therefore, their corresponding cluster 
terms, in Figure 1,  have the same integral form 

\begin{eqnarray}
I\ &=& 
\int d\vec{r}_1 \ldots d\vec{r}_s
\Gamma(\vec{r}_1,\ldots,\vec{r}_s) \,
\int d\vec{r}_i d\vec{r}_j
\gamma(\vec{r}_i,\vec{r}_j).
\label{int1}
\end{eqnarray} 

The
unlinked diagram (1a) comes from the unlinked term ($[\Gamma + \gamma]$)
in the numerator of Eq.~(\ref{gdef}) or
as a product ($[\Gamma \times \gamma]$) 
of the term ($\Gamma$) in the numerator and the term ($\gamma$)
in the denominator. The reducible diagram (1b) comes from a numerator
term. Let us now
compute for each of them the counting coefficient due to the presence of 
$\gamma(i,j)$, under the assumption 
that the diagrammatical structure $\gamma$ can be attached to any
of the $s$ vertices of $\Gamma$ in diagram (1b) (as 
in the case of Jastrow theory)
\footnote{Notice that the actual integations in $I$ lead to a factor 
$\Omega$ for each cluster integral, because of the translational 
invariance of the cluster integrands.}

\begin{eqnarray}
\left [\Gamma +\gamma \right ]_a & \rightarrow & +\frac{(N-s)(N-s-1)}
{2\Omega^2} \ , \nonumber \\
\left [\Gamma \times \gamma \right ]_a  & \rightarrow & -\frac{N(N-1)}
{2\Omega^2} \ , \nonumber \\
\left [\Gamma +\gamma \right ]_b & \rightarrow & +\frac{s(N-s)}
{\Omega^2} \ .
\label{coef}
\end{eqnarray}

Summing up the various coefficients, one 
can see that the leading terms (those in $N^2/\Omega^2$) cancel out.
Moreover, the next to leading order terms (those in $N/\Omega^2$) 
also vanish. This means that the
combined diagrammatic structures (1a) and (1b) do not contribute to the
two--body distribution function in the thermodynamic limit.
We have recovered in this example the well known property of 
the Mayer cluster expansion, that only irreducible diagrams
contribute to $g(r)$, and the reducible and unlinked diagrams 
correspond to terms of order $1/N$ or lower, which therefore 
vanish in the thermodynamic limit
\footnote{Notice that, in calculating
the normalization intergral of Eq.~(\ref{grule}), the
$1/N$ terms give a finite contribution.}.

In the case of the IPC ansatz, not all the $s$ reducible diagrams 
of type (1b) are allowed, as assumed in the previous equation. 
For instance, if a given vertex,
amongst the $s$ possible ones, has both a $B$ and a $K$ correlation
reaching it in the $\Gamma$ structure the second diagrammatical 
structure $\gamma$ cannot be attached to it (so e.g. 1b can not occur).
In this case, we do not have
anymore the cancellation of the next to leading order terms and one has 
to include  $I$ of Eq.~(\ref{int1}) (and, therefore 
both the unlinked and the reducible diagrams (1a) and (1b)) 
in the cluster expansion with the proper coefficient.

Let us now suppose that no reducible diagrams of the 
type (1b) are allowed. In this case, the leading 
coefficient of $I$, resulting from the sum of the contributions
$[\Gamma+\gamma]_a$ and $[\Gamma \times \gamma]_a$ in Eq.~(\ref{coef}),
is proportional to $N$ and given by $-s\rho/\Omega$.
The translational invariance of $\gamma$ brings in a factor $\Omega$, 
and the internal counting coefficient of $\Gamma$  another factor 
$t_{\Gamma}\times N(N-1)\ldots (N-s+1)/\Omega^s$, where $t_{\Gamma}$ is
the topological factor of $\Gamma$
\footnote{the topological factor $t_{\Gamma}$ 
is given by the inverse of the number of labels permutations 
which mantain the  same topological structure of $\Gamma$ \cite{fantoni98}}. 
Therefore the total coefficient, $\tilde{I}(\text{bosonic})$,
of the cluster term $I$ is non-zero and given by 

\begin{equation}
\tilde{I}(\text{bosonic}) = -t_{\Gamma}\times s\rho^{s+1}.
\end{equation}

Unfortunately, the summation of reducible diagrams is very difficult in 
the general case, because it requires a direct counting of terms
coming from the numerator and the denominator. 

The situation is much simpler in the Fermi case, where 
one can use the FHNC cluster expansion. With fermions of $\nu$ flavours,
because of the Pauli principle, the
various expansion terms in both the numerator and the 
denominator have coefficients, simply 
given by $t \times(-)^{s+l}\rho^s/\nu^{s-l}$, where $t$ is 
a generic topological factor, $s$ the number of vertices, 
$l$ the number of Pauli loops (counting also the one--vertex loops).
As an example, the terms with no exchanges have only
one-vertex loops, therefore $s=l$, and the coefficient 
is $t\times\rho^s$. The vertices in a
Pauli loop are connected by the two--body Fermi function
$l(k_Fr_{ij})$, given by

\begin{eqnarray}
l(x) = \frac{3}{x^3}(sin(x)-x cos(x)) \ ,
\label{slf}
\end{eqnarray}

where $x=k_Fr_{ij}$ and $k_F = (6\pi^2 \rho/\nu)^{1/3}$ 
is the Fermi momentum.

One may easily verify that, given the structure of the
coefficients of the cluster diagrams,  there is a complete 
cancellation of the
denominator with the unlinked terms of the numerator \cite{fantoni74}. 
Therefore, referring to the example of Fig. \ref{f1}, 
the corresponding \textit{fermionic} diagrams of the cluster term $I$ 
cannot be unlinked, as diagram (1a). The \textit{fermionic}
diagram corresponding to (1b) is diagram (1c), where  
Pauli one--vertex loops have been added up. The loops, however, 
do not change $I$ because
$l(k_Fr_{ii})=1$. Its global counting coefficient is 
$t_{\Gamma}\times\rho^{s+1}$. 

In addition, there are other reducible \textit{fermionic} 
diagrams, like diagram (1d), which 
have the same integral form of $I$. This is due to 
the normalization and completeness 
properties of the Fermi function $l(k_Fr_{ij})$  

\begin{eqnarray}
& & \frac{\rho}{\nu}\int d\vec{r} \, l(k_Fr) = 
\frac{\rho}{\nu}\int d\vec{r} \, l^2(k_Fr)  = 1 \ , \nonumber \\
& & \frac{\rho}{\nu}\int d\vec{r}_3 \, l(k_Fr_{12})l(k_Fr_{32}) 
= l(k_Fr_{12}) \ . 
\label{exchange}
\end{eqnarray}

It is well known that for the
Jastrow-Slater model all the reducible diagrams of the two--body
distribution function cancel out exactly, 
as for the Bose case \cite{fantoni98}.  One may 
understand such cancellation by looking at the two cluster structures
(1c) and  (1d). The coefficients of the two corresponding cluster terms are
$t_{\Gamma}\times s\rho^{s+1}$ for diagrams of type (1c) and 
$-t_{\Gamma}\times s\rho^{s+1}/\nu$ for diagrams of type (1d) 
(the separability point in both the structures (1c) and (1d) can be any
of the $s$ vertices of $\Gamma$). 
From the properties of the exchange function $l(k_Fr)$ given in
eqs. (\ref{exchange}), it follows that the two cluster terms 
cancel each other.  

However for the IPC model, diagrams (1c) are not allowed for the
same reason that (1b) is forbidden in the boson case (no B,K and 
$\gamma$ lines from a single vertex). Hence one is left in the
IPC/FHNC case with uncancelled diagrams of the (1d) type, and 
using loop integrals like \ref{exchange} ends up in the
cluster term $I$ of Eq.~(\ref{int1}) with a total coefficient 

\begin{equation}
\tilde{I}(\text{fermionic}) = -t_{\Gamma}\times s\rho^{s+1}.
\end{equation}

But this is exactly as in the case of bosonic diagrams (1a) and (1b): the
FHNC result is the same as the HNC for bosons.

From the above discussion it follows that a convenient way 
to compute the two--body distribution function
$g(r_{12})$ of Eq.~(\ref{gdef}) is to consider a notional Fermi system, 
in which the fermions have the same mass of the original bosons, 
interact with the same potential, and have a flavour degenaracy equal to
the number $N$ of particles, so that they all occupy the lowest state 
with $k=0$. We can then use the much simpler FHNC theory, 
under the constraint
$\nu=N$, in a \textit{bosonic} limit.

This $\nu=N$ limit implies $l(k_Fr_{ij})/\nu\rightarrow 1/N$. As a
consequence, all the irreducible fermionic diagrams will vanish in the 
thermodynamic limit, except those having one vertex loop only, 
which coincide
with the corresponding bosonic diagrams. 
The \textit{reducible} fermionic diagrams survive, because
integrations on the uncorrelated fermionic bonds bring $\Omega$ factors.
In the following, we will show how to calculate them in a simple 
and efficient way.

\subsection{Vertex corrections.}

In order to include reducible diagrams we use 
RFHNC theory which treats irreducible \textit{renormalized} diagrams 
characterized by vertex corrections for each of their 
vertices \cite{fantoni98}.

There are two kinds of vertex corrections

\begin{itemize}
\item \textit{bosonic vertex correction} $V_b$. This vertex correction
applies to vertices of type either $B$ or $K$, 
which are reached only by \textit{one} $B$-- or  
$K$--correlation in the underlying irreducible structure; 

\item \textit{fermionic vertex correction} $V_f$. 
This vertex correction applies 
to vertices of the type $KB$, namely those reached by \textit{both}
$K$-- and $B$--correlations in the underlying irreducible structure, and, 
therefore they can be further reached by one Pauli loop. 
  \end{itemize}

\begin{figure*}
\includegraphics{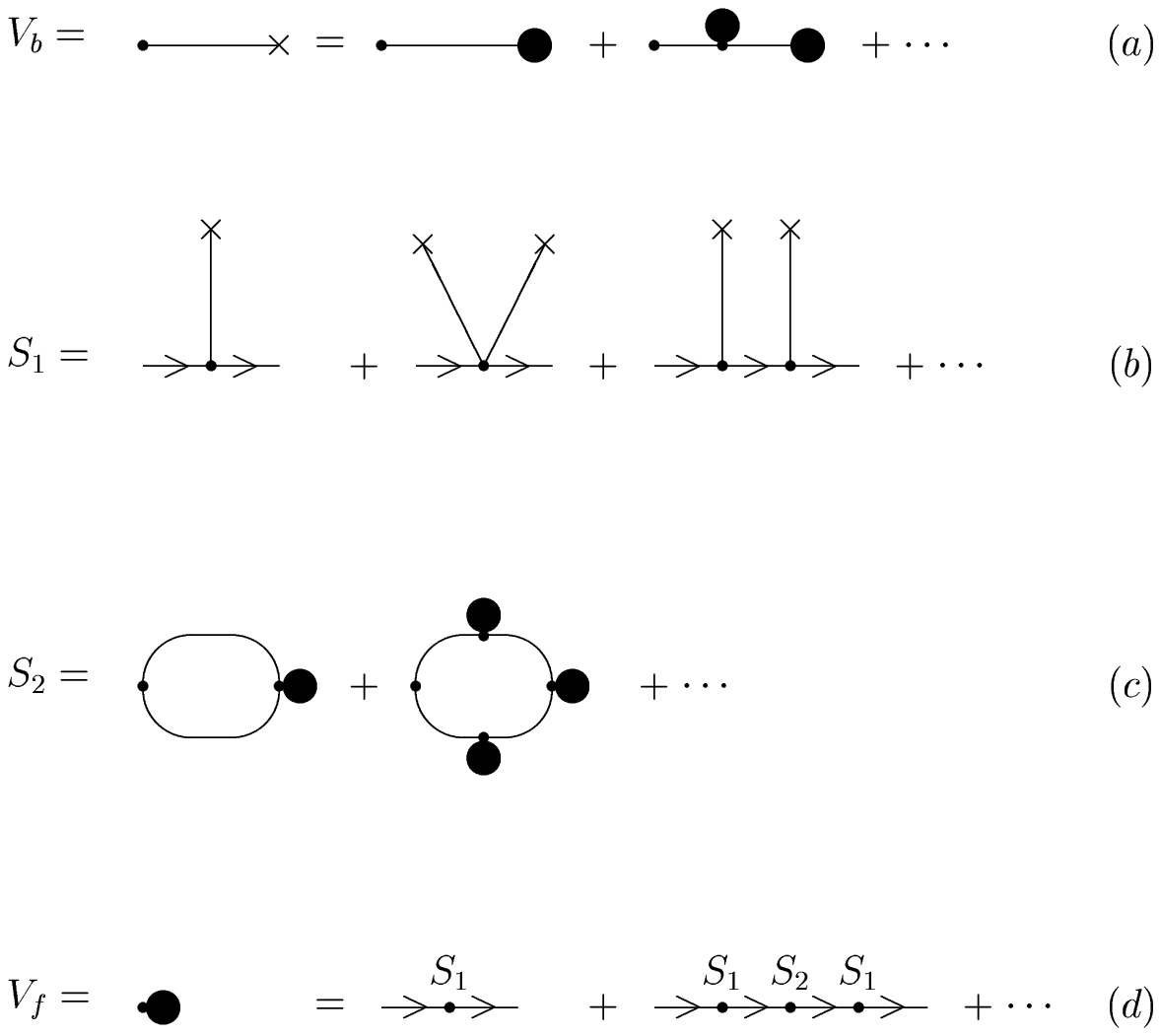}
\caption{\label{f2}
Diagrammatic equations for the vertex corrections. Equation (a) gives the
bosonic vertex correction $V_b$. Equations (b) and (c) refer to the vertex
corrections $S_1$ and $S_2$ respectively. Equation (d) exemplifies
the cluster series of the full fermionic vertex correction $V_f$. The
solid lines represent the correlation function $h(r)$, whereas the
oriented lines indicate the exchange function $l(k_Fr)$. 
}
\end{figure*}

The two $k$-independent vertex corrections $V_b$ and $V_f$ 
are inter--related and can be computed by using the
RFHNC summation technique.
The series of one--body diagrams corresponding to $V_b$ is exemplified
by Fig. \ref{f2} (a), whose sum is given by

\begin{eqnarray}
V_b = V_b(K) = V_b(B) = \frac{\tilde{h}_0 V_f}{1-
\tilde{h}_0 V_f} \ ,
\label{v1}
\end{eqnarray}

where the Fourier transform $\tilde{h}(k)$ of $h(r)$ is defined as

\begin{eqnarray}
\tilde{h}(k) = \rho\int d\vec{r} e^{\imath \vec{k}\cdot\vec{r}} h(r) \ ,
\label{ft}
\end{eqnarray}

and $\tilde{h}_0  \equiv \tilde{h}(k=0)$. 

The vertex correction $V_f$ concerns the reducible fermionic diagrams, 
which in the \textit{boson} limit ($\nu=N$) corresponds to 
\textit{unlinked} bosonic
diagrams 
(like diagram (1a)), with a coefficient given by the next to leading order
in $N$.

A series of one--body diagrams contributing to $V_f$ is displayed in 
Fig. \ref{f2} (b). The corresponding vertex correction $S_1$ is given by

\begin{eqnarray}
S_1 = 1 - 2V_b + 3V_b^2 - 4V_b^3 + \ldots = \frac{1}{(1+V_b)^2} \ .
\label{s1v}
\end{eqnarray}

Besides $S_1$, there is a second type of 
vertex correction, $S_2$,  contributing to
$V_f$. The diagrammatic representation of $S_2$ is displayed in 
Fig. \ref{f2} (c), and is, explicitly

\begin{eqnarray}
S_2 = \int \frac{d\vec{k}}{(2\pi)^3\rho} \frac{\tilde{h}^2(k) V_f}
   {1-\tilde{h}^2(k) V_f^2} \ .
\label{s2v}
   \end{eqnarray}

The fermionic vertex correction $V_f$ is made of vertices of the 
$S_1$ and $S_2$ type, as schematically shown in 
Fig. \ref{f2} (d), and it is given by

\begin{eqnarray}
V_f = \frac{S_1}{1+S_1S_2} = \frac{1}{(1+V_b)^2+S_2} \ .
\label{vfv}
   \end{eqnarray}

The vertex corrections $V_b$ and $V_f$ correctly satisfies the 
normalization 
condition for the one--body distribution function, $g_1(1)=1$.
The sum of the entire set of  one--body diagrams is given by

\begin{eqnarray}
g_1(1) = (1+V_b)^2 V_f+S_2V_f = 1 \ .
\label{g1}
   \end{eqnarray}

\subsection{Pair distribution function.}

The calculation of $g(r_{12})$ can be done by using the FHNC summation
technique for irreducible diagrams, renormalizing the vertices with
the vertex corrections $V_b$ and $V_f$, depending on the nature of the
vertex itself.
The independent pair nature of the trial function reduces the set of 
allowed cluster diagrams, and, as a consequence, the
structure of the integral equations results to be quite different from that
encountered in standard FHNC \cite{fantoni98}. The pair distribution 
function is a sum of chain and composite diagrams.

\begin{figure}
\includegraphics{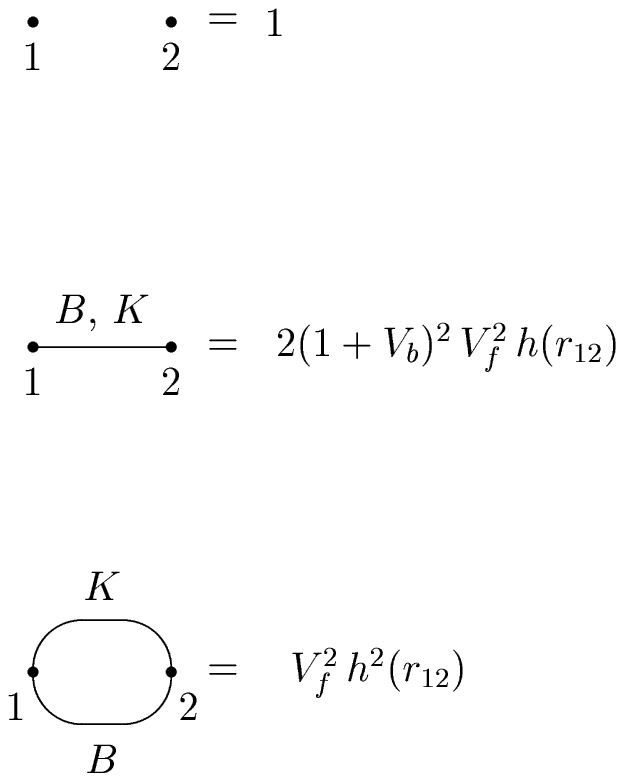}
\caption{\label{f3}
Vertex corrected two--body diagrams which contribute to the 
distribution function $g(r_{12})$,
together with the expression of the corresponding cluster terms.
The vertex corrections to the top diagrammatical structure 
$\Gamma_u(1,2)$ sum up to unity (see Eq.~(\ref{g1}))
}
\end{figure}

The lowest order diagrams contributing to $g(r_{12})$ are (i)
the ideal Bose gas (uncorrelated) diagram, $\Gamma_u(1,2)$, and
the two--body correlated diagrams, either (ii) linear,
or (iii) quadratic in $h$. They are displayed in Fig. \ref{f3}, 
together with their 
expressions, which include vertex corrections. They correspond
to the set of diagrams included in the lowest order constrained 
variational (LOCV) method \cite{pandhaLOCV}, when the Jastrow correlation 
is taken of the form $f(r)=1+V_f h(r)$ and the vertex correction 
$(1+V_b)^2 V_f$, which gives the condensate fraction 
(see Eq.~(\ref{n0})), is set equal to $1$. 

The lowest order bond, used in the chain diagrams, 
is represented by the solid line and corresponds to the
correlation function $h(r)$. After the inclusion of the vertex 
corrections, it gives the lowest order composite diagram

\begin{eqnarray}
X_0(r_{12}) = 2(1+V_b)^2\ V_f^2\ h(r_{12}) \ .
\label{x0}
\end{eqnarray}
There are four types of chain diagrams, depending on the nature ($B,K$)
of the external vertices. For symmetry reasons, $N_{BK}(1,2)=N_{KB}(2,1)
=N_{KB}(1,2)$ and  $N_{BB}(1,2)=N_{KK}(1,2)$. The allowed chain diagrams
are only those constructed with the lowest order bond and they 
are given by 

\begin{eqnarray}
\tilde{N}_{BK}(q) & = & \tilde{N}_{KB}(q) = (1+V_b)^2 V_f 
\frac{\tilde{h}^2(q) V_f^2}{1-\tilde{h}^2(q) V_f^2},
\nonumber \\
\tilde{N}_{BB}(q) & = & \tilde{N}_{KK}(q) = (1+V_b)^2 V_f
\frac{\tilde{h}^3(q) V_f^3}{1-\tilde{h}^2(q) V_f^2}. 
\label{chains}
\end{eqnarray}

The sum of all vertex corrected chain diagrams is 
 
\begin{eqnarray}
 \tilde{N}(q) &=&
2(1+V_b)^2 V_f^2 (\tilde{N}_{BK}(q)+\tilde{N}_{BB}(q)) \nonumber \\ 
&=& 2(1+V_b)^2 V_f \frac{\tilde{h}^2(q) V_f^2}{1-\tilde{h}(q) V_f} .
\label{chainsum}
   \end{eqnarray}

The vertex corrected composite diagrams contributing to $g(r_{12})$ 
are given by

\begin{eqnarray}
X(r_{12})  =  X_0(r_{12}) + X_1(r_{12}) +  X_2(r_{12}) \ ,
\end{eqnarray}

where $X_0(r)$ is given by Eq.~(\ref{x0}) and the Fourier Transforms of
$X_1(r)$ and $X_2(r)$ are

\begin{eqnarray}
\tilde{X}_1(q) & = & 
\left(\frac{\tilde{h}^2 V_f^2}{1-\tilde{h}^2 V_f^2}
|\frac{\tilde{h}^2 V_f^2}{1-\tilde{h}^2 V_f^2}
\right)_q \ , \nonumber \\
\tilde{X}_2(q) & = & 
\left(\frac{\tilde{h} V_f}{1-\tilde{h}^2 V_f^2}
    |\frac{\tilde{h} V_f}{1-\tilde{h}^2 V_f^2}
\right)_q \ ,
\label{compositeK}
\end{eqnarray}

where the convolution integral $( | )_q$ is defined by 

\begin{eqnarray}
(f|g)_q \equiv \frac{1}{(2\pi)^2\rho q}
\int_0^{\infty} dp \, p f(p)
\int_{|p-q|}^{p+q} dk \,  k g(k) \ .
\label{conv}
\end{eqnarray}

The sum of all the chain and composite diagrams gives the exact expression
of the two--body distribution function $g(r_{12})$ in the thermodynamic
limit,

\begin{eqnarray}
g(r_{12}) = 1+N(r_{12})+X(r_{12}) + \textit{O}\left(\frac{1}{N}\right) \ ,
\label{gtwo}
   \end{eqnarray}

The leading term of the $\textit{O}(1/N)$ terms is independent of the
coordinate $r_{12}$, and has the form $\alpha/N$, where the constant
$\alpha$ is given by

\begin{eqnarray}
\alpha = -1-\rho\int d\vec{r}_{12} (N(r_{12})+X(r_{12})) \ .
\label{alpha} 
   \end{eqnarray}

Notice that the constant $\alpha/N$, contributes to $S(0)$, and makes the
normalization sum rule of Eq.~(\ref{grule}) 
to be correctly satisfied, independently
on the correlation function $h(r)$ considered.
The set of cluster diagrams giving $\alpha$ is constituted by the 
renormalized two--body Pauli loop (the exchange of the
uncorrelated diagram $\Gamma_u(1,2)$ in Fig. \ref{f3}) 
plus all those corresponding to its the convolution with $g(r_{12})-1$.
Terms of the type $f(r_{12})/N$ or higher order $1/N$ terms can 
be easily calculated by using
the proposed FHNC scheme, but they are not of interest for 
the present paper.

The static structure function can be easily calculated from
Eq.~(\ref{stf}), and it results to be 

\begin{eqnarray}
S(k) & = & 1 + 2(1+V_b)^2 V_f^2 \frac{\tilde{h}(k)}{1-\tilde{h}(k) V_f} 
\nonumber \\ 
& + & 
\left(
\frac{\tilde{h}^2 V_f^2}{1-\tilde{h}^2 V_f^2}
|\frac{\tilde{h}^2 V_f^2}{1-\tilde{h}^2 V_f^2}
\right)_k  
\nonumber \\
& + &
\left(\frac{\tilde{h} V_f}{1-\tilde{h}^2 V_f^2}
    |\frac{\tilde{h} V_f}{1-\tilde{h}^2 V_f^2}
\right)_k \nonumber \\
   & + & \frac{\alpha}{\Omega}\delta(k) \ .
\label{structure}
   \end{eqnarray}

\section{Density matrices.}

We compute in this section the one-- and two--body 
density matrices  of a bulk boson system described by the IPC trial
function. They give us important information, such as the
condensate fraction, the (paired) depletion momentum distribution and the
pairing function.

The one--body density matrix $\rho_1(1,1')$ is defined as

\begin{eqnarray}
& & \rho(1,1')  = \rho(r_{11'})  =    \\
& & \Omega  \frac{\int d\vec{r}_2d\vec{r}_3\ldots 
d\vec{r}_N \Psi^{\ast}_{IPC}(1',2,\ldots,N)
   \Psi_{IPC}(1,2,\ldots,N)}{\int dR |\Psi_{IPC}(R)|^2}\ , \nonumber
\label{rho1}
\end{eqnarray}

and its related quantity, the one--particle momentum distribution is
given by

\begin{eqnarray}
n(k)  =  \langle a^{\dagger}_{\vec{k}} a_{\vec{k}}\rangle 
&=& 
\rho\int d\vec{r}_{11'} 
e^{\imath \vec{k}
\cdot\vec{r}_{11'}}
\rho(r_{11'}) \nonumber\\
&=& 
n_0 \, \delta(k) + n_d(k) \ ,
\label{nk}
  \end{eqnarray}

defining a condensate and boson-pair number distribution that can be 
independently evaluated in this IPC scheme.
Notice that, with the above definitions, 
the momentum distribution normalization is a sum rule on $n_0$ and
$n_d(k)$

\begin{eqnarray}
1 
&=&
\int \frac{d\vec{k}}{(2\pi)^3\rho} n(k)  \nonumber \\
&=&
n_0 + \int \frac{d\vec{k}}{(2\pi)^3\rho} n_d(k)  
\label{norm}
  \end{eqnarray}

\subsection{Condensate fraction and depletion contribution}

The term independent of $r_{11'}$ in the one--body density matrix, 
is the condensate fraction, and it is given by the 
product of the vertex corrections in the points $1$ and $1'$ 
\cite{fantoni78},

\begin{eqnarray}
n_0 = \lim_{r_{11'}\rightarrow\infty} \rho(r_{11'}) = 
(1+V_b)^2 V_f = \frac{(1+V_b)^2}{(1+V_b)^2+S_2} \ .
\label{n0}
  \end{eqnarray}

The coherent depletion term $n_d(k)$ is given by a diagrammatic 
series characterized by 
chains of bonds, alternatively of the $B$-- and the $K$--type,
starting with the $B$--type from point $1$ and reaching $1'$ with the
$K$--type. Each vertex has a $V_f$ correction (the external
points $1$ and $1'$ have a single vertex correction). 
The sum of this series is given by

\begin{eqnarray}
n_d(k) = \frac{\tilde{h}^2(k) V_f^2}{1-\tilde{h}^2(k) V_f^2} \ .
\label{background}
   \end{eqnarray}

The normalization sum rule (\ref{norm}) is fully satisfied

\begin{eqnarray}
\int  \frac{d\vec{k}}{(2\pi)^3\rho} n(k) = 
\frac{(1+V_b)^2}{(1+V_b)^2+S_2} + S_2V_f = 1 \ ,
\label{sumrule}
  \end{eqnarray}

and the depletion of the condensate is given by

\begin{eqnarray}
1-n_0 = S_2V_f \ .
\label{depletion}
\end{eqnarray}

\subsection{ODLRO and pairing behaviour.}

The study of the off--diagonal long range order and the calculation 
of the paring function \cite{ristig79} are based upon the 
two--body density matrix, which is defined by

\begin{eqnarray}
\label{rho2}
& &\rho_2(1,2;1',2') = \frac{N(N-1)}{\rho^2}  \\ 
&\times & \frac{\int d\vec{r}_3d\vec{r}_4\ldots d\vec{r}_N 
\Psi^{\ast}(1',2',3,\ldots,N)\Psi(1,2,3,\ldots,N)} 
{\int dR |\Psi(R)|^2} \ . \nonumber
  \end{eqnarray}

\begin{figure}
\includegraphics{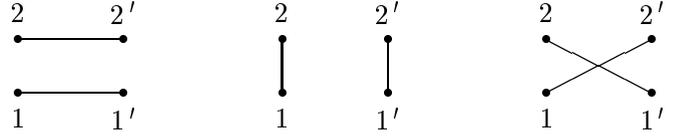}
\caption{\label{f4}
Diagrammatical structures contributing to the two--particle density
matrix. The four vertices of each structure are renormalized by their
vertex corrections, as in Fig. \ref{f3}. Each bond denote either no
correlations or a generic linked structure. The fully uncorrelated
diagram should be counted only once.
}
\end{figure}

The diagrammatical expansion of $\rho_2(1,2;1',2')$ is characterized by
the four external vertices $1$, $2$, $1'$ and $2'$. The particular 
form of the
IPC wave function excludes the possibility that more than one $h$--bond
reach any of these vertices. It follows that, in the thermodynamic
limit, the three unlinked structures in Fig. \ref{f4} are the only ones
which contribute to $\rho_2$, with the result

\begin{eqnarray}
& & \rho_2(1,2;1',2') = \rho(1,1')\rho(2,2') \nonumber \\
& & + n_0\left( n_d(r_{12'})+ n_d(r_{1'2})+P(r_{12})+P(r_{1'2'})\right)
\nonumber \\ & & + P(r_{12})P(r_{1'2'})+ n_d(r_{12'}) n_d(r_{1'2}) + 
\textit{O}\left(\frac{1}{N}\right) \ ,
\label{rho2_cluster} 
  \end{eqnarray}

where $n_d(r)$ is the reverse--Fourier transform of the depletion 
momentum distribution $n_d(k)$ of Eq.~(\ref{background}), and the
function $P(r)$, defined by 

\begin{eqnarray}
n_0 P(r) = X_0(r) + N_{BB}(r) = X_0(r) + N_{KK}(r) \ ,
  \end{eqnarray}

is the reverse--Fourier transform of the pairing function \cite{ristig79} 

\begin{eqnarray}
\widetilde{P}(k) = \frac{\tilde{h}(k) V_f}
        {1-\tilde{h}^2(k) V_f^2} = 
 -\sqrt{n_d(k)(1+n_d(k))} \ .
   \end{eqnarray}

The three diagrammatical structures of Fig. \ref{f4} refer 
to direct terms of the cluster expansion of $\rho_2$ of Eq.~(\ref{rho2}), 
namely those terms in which the coordinate $1$ in $\Psi$ corresponds 
to $1'$ in $\Psi^{\ast}$,
and $2$ to $2'$. The corresponding \textit{fermionic} diagrams have 
an exchange line joining $1$ with $1'$ and another one joining $2$ 
with $2'$, which are not drawn in the figure, and that, in the
\textit{bosonic} limit, tend to unity. The terms, in which $1$ and
$2$ are exchanged are of order $1/N$ or higher.
 
The various terms in Eq.~(\ref{rho2_cluster}) have been obtained 
as follows.

\begin{itemize} 

\item The first diagram on the left of Fig. \ref{f4} corresponds to 
the term $\rho(1,1')\rho(2,2')$ in Eq.~(\ref{rho2_cluster}), 
which also includes the diagram with no bonds giving $n_0^2$.

\item Diagrams belonging to the other two structures, 
having one bond only, contribute to terms of the type $n_0 n_d$ and 
$n_0 P$. 
Those with two bonds give rise to the \textit{pairing} and 
\textit{anti--pairing} terms $P \times P$ and $n_d \times n_d$ 
respectively. 

\item By using the HNC theory developed in section 2, one can easily
verify that the leading ${\it O}(1/N)$ term of Eq.~(\ref{rho2_cluster})
is given by  $\alpha/N$, exactly as for the two--body distribution 
function. 

\end{itemize}

Simple inspection of Eq.~(\ref{rho2_cluster}) shows
that the diagonal part of $\rho_2(1,2;1',2')$ coincides with the 
two--body distribution function $g(r_{12})$, namely

\begin{eqnarray}
\rho_2(1,2;1,2) = g(r_{12})\ .
\label{diagonal}
   \end{eqnarray}

The following independent--particle behaviour of the density 
matrix is found if the two
particles $1$ and $2$ are far away from each other
 
\begin{eqnarray}
\lim_{r_{12},r_{1'2'}\rightarrow\infty}\rho_2(1,2;1',2') = 
 \rho(1,1')\rho(2,2').
   \end{eqnarray}

However, there is another interesting limit, which is associated with 
the ODLRO. This is when $r$ and $r'$ are far away from each other

\begin{eqnarray}
& & \lim_{r_{11'},r_{22'}\rightarrow\infty}\rho_2(1,2;1',2') 
= \nonumber \\ 
& & (n_0+P(r_{12}))(n_0+P(r_{1'2'})) \ ,
\end{eqnarray}

indicating that the pairing function $P(r)$ is related to the 
superfluidity properties of the boson model system \cite{ristig79,baym69}.

The two--particle momentum distribution $n(\vec{k}_1,\vec{k}_2)=
\langle a^{\dagger}_{\vec{k}_1}a^{\dagger}_{\vec{k}_2}a_{\vec{k}_1}
a_{\vec{k}_2}\rangle$ can be computed by Fourier transforming the
two--body density matrix \cite{ristig80,fantoni81}, namely

\begin{eqnarray}
\label{nk2}
& & n_2(\vec{k}_1,\vec{k}_2) 
=  \\
& & \frac{\rho^2}{\Omega} 
\int d\vec{r}_{11'} d\vec{r}_{22'} d\vec{r}_{12} \,
e^{\imath \vec{k}_1\cdot\vec{r}_{11'}+
\imath \vec{k}_2\cdot\vec{ r}_{22'}}
\rho_2(1,2;1',2') \ . \nonumber
\end{eqnarray}

Eq. (\ref{diagonal}) garantees that
the normalization property

\begin{eqnarray}
\int \frac{d\vec{k}_1 d\vec{k}_2}{(2\pi)^6\rho^2}
n_2(\vec{ k}_1,\vec{k}_2) = (1-\frac{1}{N})\ ,
\label{normk2}
  \end{eqnarray}

is fully satisfied.

By inserting the r.h.s of  Eq.~(\ref{rho2_cluster}) into Eq.~(\ref{nk2}),
we get the result

\begin{eqnarray}
& & n_2(\vec{k}_1,\vec{k}_2)  =  n(k_1) n(k_2)+ \nonumber \\
& & \frac{1}{\Omega}\left(
\delta(\vec{k}_1)\delta(\vec{k}_2)\frac{2\rho n_0\tilde{h}_0 V_f} 
{1-\tilde{h}_0 V_f} + 
\delta(\vec{k}_1+\vec{k}_2)\widetilde{P}(k_1)\widetilde{P}(k_2)
\right.  
\nonumber \\  
 & + & 
\delta(\vec{k}_1-\vec{k}_2)n_d(k_1)n_d(k_2) 
+ \left.\delta(\vec{k}_1)\delta(\vec{k}_2)\alpha\rho 
\right) \ . 
\label{nd2k}
\end{eqnarray}

Notice that the distribution of the $(\vec{k}, -\vec{k})$ pairs 
results to be $n_d(k)^2+\widetilde{P}(k)^2$. This means that the
distribution probability $n_d^2(k)$
provided by a wave function with no ODLRO is increased by a finite 
amount given by $P^2(k)$.
A similar feature is obtained for the $(\vec{k},\vec{k})$ pairs, whose
distribution probability is increased by $n_d^2(k)$. 

\section{Energy expectation value and Euler equation.}

The energy expectation value is defined as 

\begin{eqnarray}
E[h]= \frac{\langle V\rangle}{N} + \frac{\langle T\rangle}{N} \ .
\label{eev}
   \end{eqnarray}

The potential energy $\langle V \rangle/N$ is given by

\begin{eqnarray}
\frac{\langle V\rangle}{N} & = & \frac{1}{2}\rho\int d\vec{r} v(r) g(r)  
\\
& = & \frac{1}{2}\rho\int \frac{d\vec{q}}{(2\pi)^3\rho} 
\tilde{v}(q) (S(q)-1) + \frac{1}{2}\rho\int d\vec{r} v(r) \ , \nonumber
\label{pot}
   \end{eqnarray}

which can be computed by using either the expression for the pair function
$g(r)$ given in Eq.~(\ref{gtwo}) or the static structure function of 
Eq.~(\ref{structure}). Both $g(r)$ and $S(q)$ depend upon $h(r)$, 
therefore $\langle V \rangle /N$ is a functional of $h(r)$.

The  kinetic energy expectation value is defined as 

\begin{eqnarray}
\frac{\langle T \rangle}{N} = -\frac{\hbar^2}{2m}
\frac{\langle \Psi|\nabla^2_1 |\Psi\rangle}
{\langle\Psi|\Psi\rangle}\ ,
\label{tr}
   \end{eqnarray}

The action of $\nabla^2_1$ on $\langle R|\Psi\rangle$ 
does not lead to three--body operators
as in the Jastrow case, since

\begin{eqnarray}
-\frac{\hbar^2}{2m} \nabla^2_1 \Psi(R) &=& 
\sum_{i\ne 1} t(r_{1i})  \nonumber \\
&+& 
\sum_{(i\ne 1)\neq (l\leq m)}  t(r_{1i})  h(r_{lm}) + \ldots\ 
\label{tpsi}
   \end{eqnarray}

where

\begin{eqnarray}
t(r) & = & -\frac{\hbar^2}{2m} \frac{1}{r} 
(h(r) \, r )'' \ , \nonumber \\
\tilde{t}(q) & = & \epsilon(q)\tilde{h}(q)\ ,
\label{t0}
   \end{eqnarray}

and $\epsilon(q) = \hbar^2q^2/(2 m)$ is the uncorrelated single
particle energy.  

The cluster expansion of $\langle T \rangle /N$ is similar to that of 
the pair distribution function. The two--body diagram, with only one 
correlation line ($t(r_{12})$) between the two external points $1$ 
and $2$, gives no contribution under $r_{12}$--integration. Therefore, 
one is left with the two--body correlated diagram, having both 
$t(r_{12})$ and $h(r_{12})$ as correlation lines, plus the composite 
diagrams given by $t(r_{12})$ dressed by  $N_{LL}(r_{12})$ chains, 
leading to the  following expression

\begin{eqnarray}
\frac{\langle T\rangle}{N} = \int \frac{d \vec{q}}{(2\pi)^3\rho}
\frac{\tilde{t}(q)\tilde{h}(q) V_f^2}{1-\tilde{h}^2(q) V_f^2}
=\int \frac{d \vec{q}}{(2\pi)^3\rho} 
\epsilon(q) n(q) \ .
\label{Texpression}
\end{eqnarray}

The above equation proves that the kinetic energy sum rule is exactly
satisfied in the IPC model with the FHNC treatment.

The total energy per particle $E = \langle T\rangle /N + 
\langle V\rangle /N $ 
can be expressed in terms of the momentum distribution,

\begin{eqnarray}
 E[n_d] &=&  \int \frac{d \vec{q}}{(2\pi)^3 \rho} \epsilon(q) n_d(q)
+ \frac{1}{2}\tilde{v}(0)  \nonumber \\
& + & n_o \int \frac{d \vec{q}}{(2\pi)^3 \rho}
 \Big( n_d(q) + \widetilde{P}(q) \Big) \tilde{v}(q) \nonumber \\ 
& + &
 \frac{1}{2} \int \frac{d \vec{q}}{(2\pi)^3 \rho} \tilde{v}(q)
 \left( n_d | n_d \right)_q \nonumber \\
 &+& \frac{1}{2} \int \frac{d \vec{q}}{(2\pi)^3 \rho} \tilde{v}(q)
 \left( \widetilde{P}|\widetilde{P} \right )_q \ .
   \label{energy}
  \end{eqnarray}

\subsection{Applying the Ritz variational principle.}

The Euler equation is obtained by applying the Ritz variational 
principle to the energy per particle $E[h]$.

\begin{eqnarray}
\frac{\delta}{\delta \tilde{h}(q)} E[h] =0 \ ,
\label{eulerq}
   \end{eqnarray}

which gives rise to an integro--differential equation 
having $\tilde{h}(q)$ as functional variable, derived in the Appendix.

Since the total energy can be expressed in terms of the
background momentum distribution $n_d(q)$ only, as shown
in Eq.~(\ref{energy}), then Eq.~(\ref{eulerq}) can be written as

\begin{eqnarray}
 \frac{\delta E}{\delta \tilde{h}(q)} = 
 \int dk \frac{\delta E}{\delta n_d(k)} 
 \frac{\delta n_d(k)}{\delta \tilde{h}(q)} = 0 \ ,
\label{euler_n_to_h}
\end{eqnarray}

and from Eq.~(\ref{energy})

\begin{eqnarray}
\frac{\delta E[n_d]}{\delta n_d(q)} & = & \frac{q^2}{2 \pi^2 \rho} 
\left(
\epsilon(q) \right. \nonumber \\
 &+&  \left. n_0 \tilde{v}(q) \left(1-\frac{2 n_d(q)+1}
{2 \sqrt{n_d(q)(1+n_d(q))}}\right) \right.  \nonumber \\
&+ & \left. \left(\tilde{v}|n_d\right)_q
- \frac{2 n_d(q)+1}{2 \sqrt{n_d(q)(1+n_d(q))}} 
\left(\tilde{v}|\widetilde{P}\right)_q   \right. \nonumber \\
& - & \left. \int \frac{d\vec{k}}{(2 \pi)^3 \rho} 
\tilde{v}(k) \Big( n_d(k)+\widetilde{P}(k) \Big) \right) 
\label{E_vs_n}
\end{eqnarray}

and the variation of $n_d(q)$ with respect to $\tilde{h}(k)$ is 
given by

\begin{eqnarray}
\frac{\delta n_d(k)}{\delta \tilde{h}(q)} = 2 V_f F(k) \delta (q-k) + 
\frac{\delta V_f}{\delta \tilde{h}(q)} 2 \tilde{h}(k) F(k) \ ,
   \label{n_vs_h}
 \end{eqnarray}

where the explicit expressions of the 
functions $F(k)$ and $\delta V_f/\delta \tilde{h}(q)$ 
can be found in the Appendix (see eqs. (\ref{Ccorr}) and (\ref{varver}) 
respectively). Inserting Eq.~(\ref{n_vs_h}) 
into Eq.~(\ref{euler_n_to_h}), the Eq.~(\ref{E_vs_n}) can be written as

\begin{eqnarray}
\frac{\delta E[n_d]}{\delta n_d(k)} = 
\frac{k^2}{2 \pi^2 \rho} A[n_d] \ ,
\label{solut}
 \end{eqnarray}

where the $k$--independent term $A[n_d]$ is given by

\begin{eqnarray}
&&  A[n_d] = 
 \frac{2}{V_f^2 (a_1+C)} 
\nonumber \\
&&
\int_0^{\infty} dk n_d(k)\Big(1+n_d(k)\Big) 
 \times \frac{\delta E[n_d]}{\delta n_d(k)} \ ,
\label{euler_in_n}
 \end{eqnarray}

with the constants $a_1$ and $C$ defined in eqs. (\ref{varversol}) and
(\ref{Ccorr}) respectively. Inserting the explicit expression of
$\delta E[n_d]/\delta n_d(k)$, given 
in Eq.~(\ref{E_vs_n}) into eqs. (\ref{solut})
and (\ref{euler_in_n}), we get the following system of 
integro--differential equations

\begin{eqnarray}
& & \frac{2A}{V_f^2 (a_1+C)} \int \frac{d \vec{k}}{(2 \pi)^3 \rho} 
  n_d(k) \Big( 1 + n_d(k) \Big) = A \ ,
\label{constraint}
\end{eqnarray}

where

\begin{eqnarray} 
 A[n_d(q)] &=& 
\epsilon(q) + n_0 \tilde{v}(q) \left(1-\frac{2 n_d(q)+1}
{2 \sqrt{n_d(q)(1+n_d(q))}}\right) \nonumber \\
&+& \left(\tilde{v}|n_d\right)_q 
 -\frac{2 n_d(q)+1}{2 \sqrt{n_d(q)(1+n_d(q))}} 
\left(\tilde{v}|\widetilde{P}\right)_q  
\nonumber \\ 
&-& \int \frac{d\vec{k}}{(2 \pi)^3 \rho} 
\tilde{v}(k) \left( n_d(k)+\widetilde{P}(k) \right) 
\label{eul_res}
\end{eqnarray}

Going back to Eq.~(\ref{euler_n_to_h}) we see that one optimal solution
is $A=0$. However, in Eq.~(\ref{A_coeff_s}) below we see that A may be 
non zero in order for the boson-pair number distribution 
$n_d(k) \sim 1/k$ so the compressibility sum rule can be obeyed. 
Hence we consider only this solution.

\subsection{The IPC Euler equation}

It is well known that the depletion component of the momentum
distribution, $n_d(q)$ has a $1/q$ singularity in the long
wavelength limit \cite{nozieres64}. The behavior $n_d(q)\rightarrow 
n_0 mc/(2q)$ at small $q$, where $c$ is the sound velocity, 
uniquely depends upon the existence of a condensate and of a finite
value of the compressibility \cite{nozieres64,stringari95}.

According to this important property, the long wavelength behavior of
$n_d(q)$ can be written as

\begin{eqnarray}
 n_d(q) \to \frac{d_{-1}}{q} + d_0 + d_1 q + \ldots
\qquad \text{as} \qquad q \to 0^+\ ,
\label{singu_n}
 \end{eqnarray}

and, consequently, $\tilde{h}(q)V_f\rightarrow -1+q/(2d_{-1})$ 
as $q\rightarrow 0$. Using the above long wavelength behaviors, and
developing the r.h.s. of Eq.~(\ref{eul_res}) around $q=0$, one gets the
solution

\begin{eqnarray}
     A = - 2 \int \frac{d \vec{k}}{(2 \pi)^3 \rho} 
     \tilde{v}(k) \widetilde{P}(k) \ .
   \label{A_coeff_s}
  \end{eqnarray}

By substituting Eq.~(\ref{A_coeff_s}) in Eq.~(\ref{eul_res}) one gets 
the following IPC  Euler equation

\begin{eqnarray}
& & 
 \epsilon(q) + n_0 \tilde{v}(q) \left(1-\frac{2 n_d(q)+1}
{2 \sqrt{n_d(q)(1+n_d(q))}}\right) \nonumber \\
& &+  \left(\tilde{v}|n\right)_q 
-  \frac{2 n_d(q)+1}{2 \sqrt{n_d(q)(1+n_d(q))}} 
\left(\tilde{v}|\widetilde{P}\right)_q    \nonumber \\
& &-\int \frac{d\vec{k}}{(2 \pi)^3 \rho} 
\tilde{v}(k) \left( n_d(k)-\widetilde{P}(k) \right)  = 0 \ ,
\label{euler_new}
\end{eqnarray}

which is the a central result of the present paper.
It is important to notice that the solution of Eq.~(\ref{euler_new}),
automatically satisfies Eq.~(\ref{constraint}). Substituting the
expressions of $a_1$ and $C$ into Eq.~(\ref{constraint}), and 
using the long wavelength behavior $\tilde{h}_0V_f=-1$, one gets 

\begin{eqnarray}
\int \frac{d \vec{k}}{(2 \pi)^3 \rho} n_d(k) = 1 + 2 V_f 
\frac{\tilde{h}_0 V_f}{(1-\tilde{h}_0 V_f)^3} = 1-\frac{1}{4}V_f \ ,
 \label{solut_2_2}
  \end{eqnarray}

which can be used to express the vertex correction $V_f$ in terms of
the condensate fraction $n_0$,

\begin{eqnarray}
V_f = 4 \left( 1 - \int \frac{d \vec{k}}{(2 \pi)^3 \rho} n_d(k) \right)
   = 4 n_0 \ .
 \end{eqnarray}

The same results, $V_b=1/2$ and $V_f=4n_0$
follow from the vertex correction equations (\ref{v1}) 
and (\ref{vfv})
and the depletion of the condensate given in Eq.~(\ref{depletion}).

Therefore, the IPC Euler equation (\ref{euler_new}) is consistent with
the Gavoret--Nozi\'eres singularity for the long--wavelength 
behavior of the momentum distribution of a quantum Bose fluid. 

The long wavelength behavior of the static structure function $S(q)$,
resulting from the solution of the Euler equation (\ref{euler_new}) is
given by

\begin{eqnarray}
S(q) \ &\to &  S(0^+) + q \frac{dS(0^+)}{dq}
\qquad \text{as} \qquad q \to 0^+ \ , \nonumber \\
S(0^+) &=& 2 \int \frac{d\vec{k}}{(2\pi)^3\rho} P^2(k) \ > \ 0 \ .
\label{lwsq}
\end{eqnarray}

Notice that for the ideal Bose gas $S(q) = S(0^+) = 1$, whereas for
the Jastrow ansatz, as well as for the exact eigenfunction, 
$S(0^+) = 0$. Therefore, pair correlations alone are not sufficient 
to bring $S(0^+)$ from one to zero. However, we will show that for the 
IPC model, and its Bogoliubov limit reasonable potentials yield 
$S(0^+) \sim 0$. Furthemore 
$\left. dS(q^+)/dq \, \right \vert_{q=0} \sim 1/(2mc)$, namely the 
long wavelength behavior of $S(q)$ is very close to that required by 
sum rules \cite{stringari95}.
 
\subsection{Bogoliubov approximation}

It is interesting to understand under which approximations 
our HNC theory recovers the Bogoliubov-type result. Since the
sum of the chain diagrams corresponds to RPA theory, we expect that
neglecting the composite diagrams $X_1$ and $X_2$ 
in the  HNC summation lead to the
Bogoliubov approximation \cite{nozieres99}. Under this approximation,
and setting the condensation fraction $n_0=1$, the
total energy per particle of Eq.~(\ref{energy}) is given by 
\cite{huang,pines59}

\begin{widetext}
\begin{eqnarray}
E_{chain}[n_d] 
&=& E_{un}[n_d]  
\label{energychain}
\\
&+& 
\int \frac{d \vec{q}}{(2\pi)^3 \rho}
\left( \epsilon(q) n_d(q) +  \tilde{v}(q)
      \left( 
            n_d(q) - \sqrt{n_d(q)+n_d(q)^2} 
            \right)
\right) \ , \nonumber
\end{eqnarray}
\end{widetext}

where

\begin{eqnarray}
E_{un}[n_d] = \frac{1}{2}\tilde{v}(0) \ ,
\label{energyunc} 
\end{eqnarray}

is the energy upperbound obtained with the uncorrelated variational
function $\Psi(R)=1$.

Using Eq.~(\ref{energychain}) for the total energy, 
the Euler equation becomes

 \begin{eqnarray}
 \epsilon(q) + \tilde{v}(q)\left(1-
\frac{2 n_d(q)+1}{2 \sqrt{n_d(q)+n_d^2(q)}}\right) = 0\ ,
 \end{eqnarray}

which yields the Bogoliubov solution for the momentum distribution

\begin{eqnarray}
  n_{Bog}(q) &=& \frac{1}{2\omega(q)}\left(\epsilon(q)+\tilde{v}(q)-
\omega(q)\right) \ , \nonumber \\
\omega^2(q) &\equiv& \epsilon^2(q)+2 \epsilon(q)\tilde{v}(q) \ . 
  \end{eqnarray}

$n_{Bog}(q)$ goes as $\sqrt{E_{un}}/(2q)$ in the long wavelength
limit. This Bogoliubov momentum distribution function  has been used as
starting point in the numerical solution of the IPC Euler equation 
(\ref{euler_new}). Inserting $n_{Bog}(q)$  back into 
Eq.~(\ref{energychain})
one recovers the Bogoliubov estimate for the total energy per particle

\begin{eqnarray}
E_{Bog} = \frac{1}{2} \tilde{v}(0)
&+& \int \frac{d \vec{q}}{(2\pi)^3 \rho}
\left(n_{Bog}(q)\left(\epsilon(q)+
\tilde{v}(q)\right) \right. \nonumber \\
&-& \left. \tilde{v}(q)\sqrt{n_{Bog}(q)+n_{Bog}^2(q)}\right) . 
\label{energybog}
\end{eqnarray}

The static structure function $S_{Bog}(q)$ has a long wavelength
behavior characterized by $S(0^+) = 1-n_0$ and 
$dS(q)/dq \vert_{q=0^+}= n_0/(2\sqrt{E_{un}})$.

The Bogoliubov approximation, and therefore our independent pair model, 
is best for those interactions such that the scattering length $a$ 
coincides with its Born approximation

\begin{eqnarray}
a_{\text{Born}} = \frac{m}{4\pi\hbar^2}\int d\vec{r} v(r) 
= \frac{m}{4\pi\rho\hbar^2}\tilde{v}(0) \ .
\end{eqnarray}

Notice that 
$a=a_{\text{Born}}$ in the case of a $\delta$--force interaction 
of the  type $v(r) =  4\pi\hbar^2 a\delta(\vec{r})/m$ \cite{huang}.
For such an interaction, the Bogoliubov
approximation coincides with the 
Lee and Yang \cite{lee57} low--density expression

\begin{eqnarray}
E_{LY} = \left(\frac{\hbar^2}{2ma^2}\right) 4\pi\rho a^3\left[1+
\frac{128}{15} \sqrt{\frac{\rho a^3}{\pi}}\right] \ ,
\label{energyLY}
\end{eqnarray}

and the momentum distribution goes as $q^{-4}$ at large $q$.

\section{Results}

In this section we present and discuss the results obtained for the
equation of state, the two--body distribution function, the 
momentum distribution and the pairing function of a 
weakly interacting Bose system within the
independent pair model, in correspondence with two different class of
potentials.

For low density systems, like atomic vapours, of interest in the 
BEC physics, the only quantity which is considered to characterize 
the interaction is the scattering length $a$, namely its strength. 
For such systems, $a\ll r_0$, where
$r_0=(3/(4\pi\rho))^{1/3}$ is the average distance between the particles.
Recently, the possibility of creating Bose systems with relatively large
values of $\rho a^3$ has been opened up \cite{cornish00,cowell01}. 
For $\rho a^3 > 10^{-4}$ the shape of the interaction 
becomes important for quantities such as 
the energy per particle and the condensate fraction \cite{giorgini99}.

In this paper we consider the properties of a Bose gas for $\rho a ^3$
in the range $10^{-6}\ - \ 10^{-1}$, or, equivalently for 
$0.016 \leq (a/r_0)\leq 0.75$. The pseudo--potentials $v(r)$ used  are
purely repulsive and have a range of 
interaction which goes from $\sim 0.1 r_0$ to $\sim 10 r_0$. 
We have considered two different choices for $v(r)$ 

\begin{description}
\item[(i)] Repulsive gaussian (G) potential defined by

\begin{eqnarray}
v(r) = v_0 \, e^{-\frac{r^2}{2\sigma^2} } \ ,
\end{eqnarray}

where $\sigma$ is the width of the gaussian. Table \ref{table1} 
gives the values of the parameters $v_0$ and $\sigma$ 
of three different choices of gaussian potentials, labelled 
with G1, G2 and G3. 
 
\item[(ii)] Soft--sphere (SS) potential defined by

\begin{equation}
v(r) =  \left\{ \begin{array}{ll}
           v_0   & \mbox{if $r\leq R$} \ ,\nonumber \\
           0     & \mbox{otherwise} \ ,
           \end{array}
           \right.  
\end{equation}
where we have taken $R=5a$ and $R=10a$ as in \cite{giorgini99}. 
The values of the parameters are reported in Table \ref{table2}.

\end{description}

\begin{table}
\caption{\label{table1}
Gaussian potentials used in the calculations of the weakly 
interacting Bose system. 
The strength $v_0$ and the energy $E_{un}$ of the uncorrelated system 
($\Psi(R)=1$) are given in units
$\hbar^2/(2ma^2)$. The width $\sigma$ and the effective range $r_{eff}$
are in units of the scattering length $a$. 
}
\begin{ruledtabular}
\begin{tabular}{cccccc}
Potential &  $v_0$  &  $\sigma$  & $a/a_B$ & $r_{eff}$ & $E_{un}/\rho$ \\
\hline
 G1   & $1.1778\times 10^{-3}$ & $11.256$  & $0.95$  & $-230.089$& $13.228$ \\
 G2   & $1.0028\times 10^{-2}$ & $5.6127$  & $0.90$  & $-51.369$ & $13.963$ \\
 G3   & $9.1978\times 10^{-2}$ & $2.7887$  & $0.80$  & $-9.759$  & $15.708$ \\
\end{tabular}
\end{ruledtabular}
\end{table}

\begin{table}
\caption{\label{table2}
Soft--sphere potentials used in the calculations of the weakly 
interacting Bose system and taken from Ref. \cite{giorgini99}. 
The strength $v_0$ and the energy $E_{un}$ of the uncorrelated system
($\Psi(R)=1$) are  given in units
$\hbar^2/(2ma^2)$. The range parameter $R$ and the effective range $r_{eff}$
are in units of the scattering length $a$. 
}
\begin{ruledtabular}
\begin{tabular}{cccccc}
Potential & $v_0$ & $R$ & $a/a_B$ & $r_{eff}$ & $E_{un}/\rho$ \\
\hline
 SS10  & $6.8167\times 10^{-3}$ & $10$  & $0.880$ & $-29.936$& $14.280$ \\
 SS5   & $6.3086\times 10^{-2}$ & $5$   & $0.761$ & $-4.9637$& $16.513$ \\
\end{tabular}
\end{ruledtabular}
\end{table}

Tables \ref{table1}  and \ref{table2} give also the 
effective range parameter $r_{eff}$, the ratio $a/a_{\text{Born}}$
and the energy upperbound $E_{un}$ 
provided by the uncorrelated (ideal) Bose gas.
The scattering length $a$ has been calculated by using a standard
procedures \cite{joachain79}.
Notice in the case of a delta--type potential, when $a/a_{\text{Born}}=1$
and $E_{Bog} = E_{LY}$, the upperbound $E_{un}$ coincides with the 
lowest order term of the Lee and Yang expansion, 
$E_{LY0} = (\hbar^2/(2ma^2))4\pi\rho a^3$. $E_{LY0}$ has been proved to be
an energy lower bound in the limit of $\rho a^3 \rightarrow 0$ and 
under general conditions of the interaction, which are met by
the potentials considered here \cite{lieb98}.

\begin{figure*}
\includegraphics{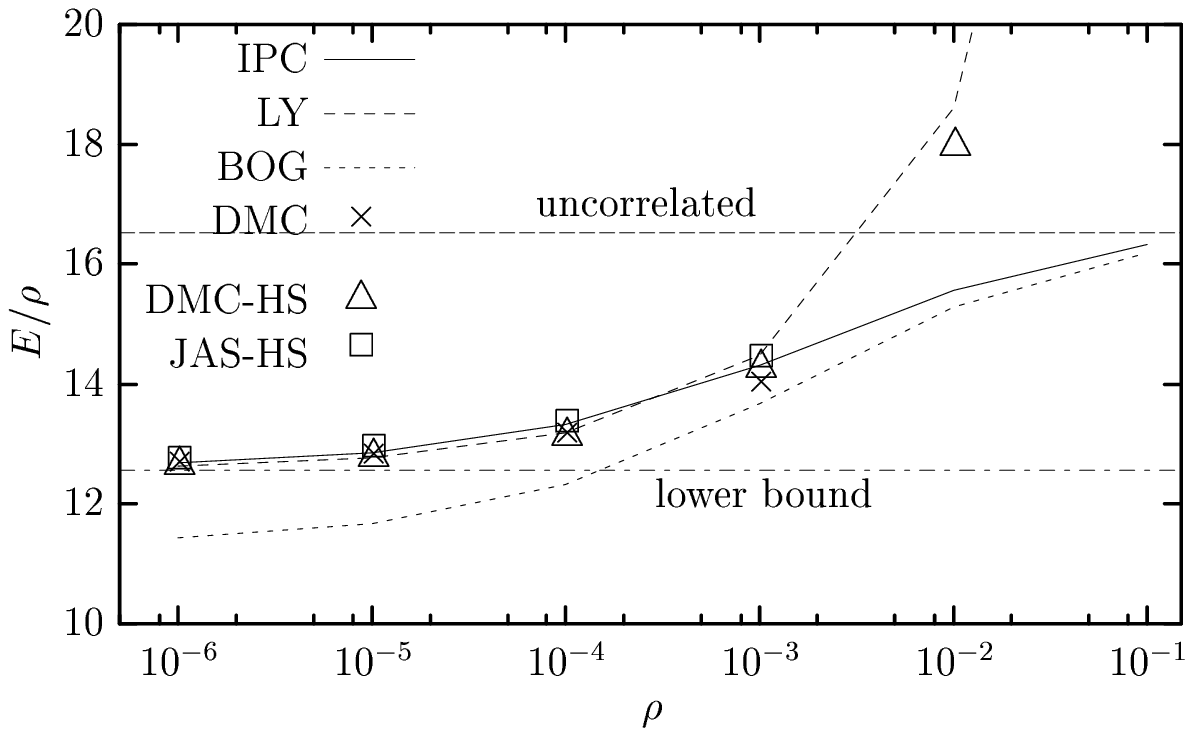}
\caption{\label{f5}
Energy per particle or equation of state of a weakly interacting Bose
gas for the soft--sphere potential SS5. 
The IPC results (solid line) are compared with the low density estimates 
$E_{LY}$ (dashed line), those coming from the Bogoliubov approximation 
(dotted line), $E_{Bog}$, and the DMC
calculations of Ref. \cite{giorgini99} (crosses). 
The DMC \cite{giorgini99} (triangles) and Jastrow \cite{fabrocini99} 
(squares) results obtained in correspondence with 
the hard--sphere potential are also reported for completeness. 
The energies are in units of $\hbar^2/(2ma^2)$, 
and the density $\rho$ in units of $a^{-3}$.
}
\end{figure*}

\begin{figure*}
\includegraphics{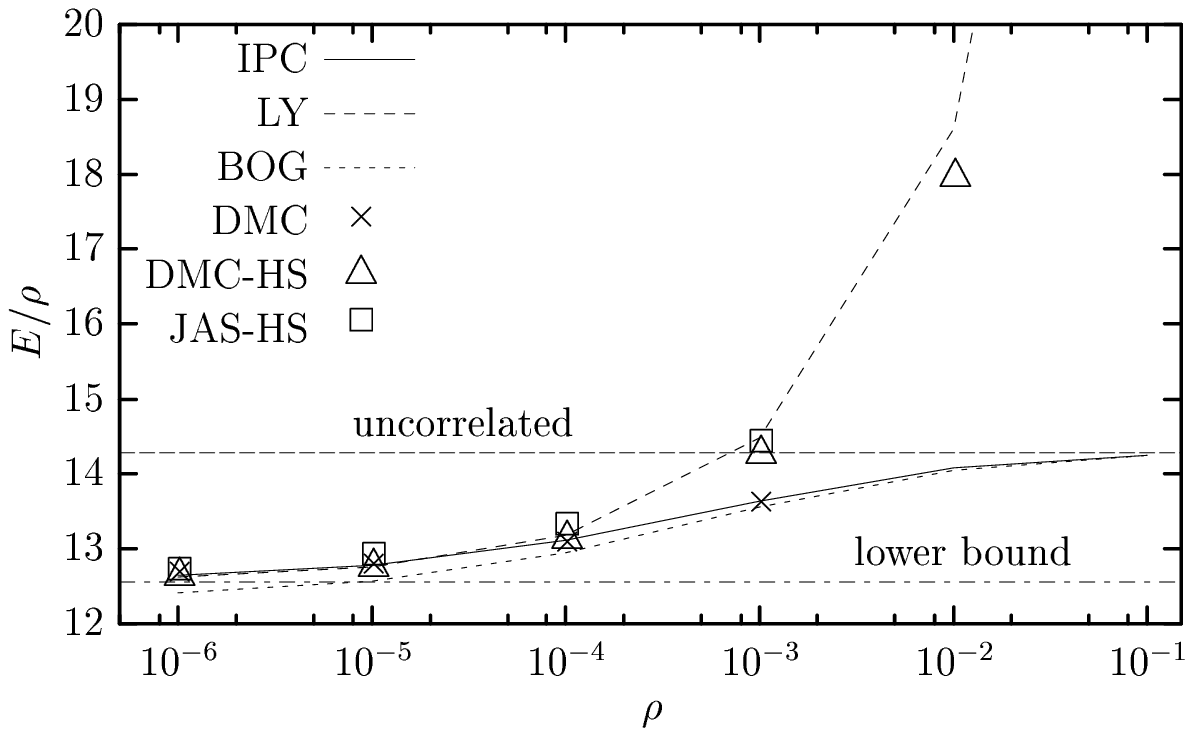}
\caption{\label{f6}
Energy per particle or equation of state of a weakly interacting Bose
gas for the soft--sphere potential SS10. 
The IPC results (solid line) are compared with the low density estimates
$E_{LY}$ (dashed line), those coming from the Bogoliubov approximation 
(dotted line), $E_{Bog}$, and the DMC
calculations of Ref. \cite{giorgini99} (crosses). 
The DMC \cite{giorgini99} (triangles) and Jastrow 
\cite{fabrocini99} (squares) results obtained in correspndence with 
the hard--sphere potential are also reported for completeness.
The energies are in units of $\hbar^2/(2ma^2)$,
and the density $\rho$ in units of $a^{-3}$.
}
\end{figure*}

The Euler equation (\ref{euler_new}) has been solved numerically by means 
of a self--consistent iterative procedure. Few iterations were sufficient
to reach convergence for all the cases considered.
The \textit{optimal} $n_d(q)$ has 
been found to have the $n_0 mc/(2q)$ behavior in the long wavelength limit.
The results for the IPC energy per particle of Eq.~(\ref{energy}) 
are displayed in Table \ref{table3} in correspondence with 
the five potentials considered, and in Figs. \ref{f5} and \ref{f6} for
the soft--sphere potentials.  $E_{IPC}$ can be compared with 
the upperbounds provided by $E_{un}$ of Eq.~(\ref{energyunc}), 
given in Tables \ref{table1} and \ref{table2} and the
estimates from the low density expansion, $E_{LY}$, of 
Eq.~(\ref{energyLY}), which are independent on the shape of the potential. 
Figs. \ref{f5} and \ref{f6} compare the IPC results  
with $E_{Bog}$ of Eq.~(\ref{energybog}) and with 
the available DMC calculations \cite{giorgini99}. The figures display  
also the DMC \cite{giorgini99} and Jastrow \cite{fabrocini99} results 
for the  hard--sphere potential.

\begin{table}
\caption[]{\label{table3}
Energy per particle of the IPC model divided by the density, 
$E_{IPC}/\rho$, at various densities, for the five potentials 
considered. The IPC upperbounds are 
compared with the low--density expansion estimates of 
Eq.~(\ref{energyLY}).
The energies are given in units of $\hbar^2/(2ma^2)$ 
and the density $\rho$ in $a^{-3}$. In these units 
$E_{LY0}/\rho = 4\pi = 12.556$ 

}
\begin{ruledtabular}
\begin{tabular}{llccccc}
$\rho$ & $E_{LY}/\rho$ & $G1$ & $G2$ & $G3$ & $SS10$ & $SS5$  \\
\hline
 $10^{-6}$ & $12.63$ & $12.63$& $12.63$ & $12.64$ & $12.65$ & $12.68$ \\
 $10^{-5}$ & $12.76$ & $12.73$& $12.76$ & $12.80$ & $12.78$ & $12.85$ \\
 $10^{-4}$ & $13.17$ & $12.92$& $13.05$ & $13.22$ & $13.11$ & $13.31$ \\
 $10^{-3}$ & $14.48$ & $13.11$& $13.48$ & $14.06$ & $13.64$ & $14.31$ \\
 $10^{-2}$ & $18.62$ & $13.20$& $13.81$ & $15.01$ & $14.08$ & $15.56$ \\
 $10^{-1}$ & $31.70$ & $13.22$& $13.93$ & $15.55$ & $14.24$ & $16.33$ \\
\end{tabular}
\end{ruledtabular}
\end{table}

The above results deserve the following comments

\begin{description}

\item[(i)] the IPC upperbounds are above the Bogoliubov estimates for all 
the pseudopotentials and the density values considered. Therefore, the
terms of the hamiltonian neglected in Bogoliubov theory give a repulsive
contribution.
Notice that, contrary to $E_{IPC}$, $E_{Bog}$ is not an energy upperbound. 
The relative differences between the
IPC and Bogoliubov results, $(E_{IPC}-E_{Bog})/E_{Bog}$, are reported on
Table \ref{table4}. One can see that $E_{IPC}$ and $E_{Bog}$ get closer 
and closer when either $a/a_{\text{Born}}\rightarrow 1$ or for increasing 
values of $\rho a^3$. In these two limits they both approach $E_{un}$ 
from below.

\item[(ii)] the low density expression of Eq.~(\ref{energyLY}) 
starts failing already at $\rho a^3
\sim 10^{-3}$. At this density value and above it $E_{LY}$ is
always larger than $E_{IPC}$. The next to leading order in the low density
expansion, giving a term in $\rho^2 \log(\rho)$ \cite{wu59}, improves the
total energy estimates up to $\rho a^3 \sim 10^{-4}$ (adding this correction
term to $E_{LY}$ the low density expansion agrees with DMC \cite{giorgini99}
within three digits).  However, it gives too much attraction at higher 
density values. For instance, in the units of Fig. \ref{f5}, the correction
at $\rho a^3 = 10^{-3}$ is $-1.706$, and at $\rho a^3 = 10^{-2}$ 
is $-11.37$. 

\item[(iii)] there is an overall agreement of 
the IPC upperbounds with the available 
DMC \cite{giorgini99} results, 
even for the SS5 potential, which is the one with the 
shorter range of interaction. Contrary to $E_{LY}$ and to $E_{Bog}$, 
both the IPC and DMC results are 
sandwiched between $E_{un}$ and the lower bound $E_{LY0}$ in the full
range of density values considered.

\item[(iv)] the shape of the potential starts becoming important already
at $\rho a^3 \sim 10^{-4}$. Therefore, in the region of large $a$, the 
scattering length is not anymore sufficient to
determine the interaction.
 
\end{description}

\begin{table}
\caption[]{\label{table4}
Comparison between the IPC model and the Bogoliubov theory.
The table reports the relative difference $\frac{E_{IPC}-E_{Bog}}
{E_{Bog}}\times 10^3 $ for the five potentials considered.
The density is given in units of $a^{-3}$. 
}
\begin{ruledtabular}
\begin{tabular}{lccccc}
$\rho$     & $G1$    & $G2$   & $G3$    & $SS10$  & $SS5$  \\
\hline
 $10^{-6}$ & $2.589$& $12.34$ & $68.05$ & $19.11$ & $109.6$ \\
 $10^{-5}$ & $2.053$& $10.80$ & $62.62$ & $17.03$ & $101.4$ \\
 $10^{-4}$ & $1.143$& $7.541$ & $49.54$ & $12.51$ & $81.32$ \\
 $10^{-3}$ & $0.355$& $3.353$ & $28.19$ & $6.337$ & $47.82$ \\
 $10^{-2}$ & $0.063$& $0.836$ & $9.934$ & $2.451$ & $18.43$ \\
 $10^{-1}$ & $0.010$& $0.163$ & $2.713$ & $0.276$ & $9.003$ \\
\end{tabular}
\end{ruledtabular}
\end{table}

\begin{figure*}
\includegraphics{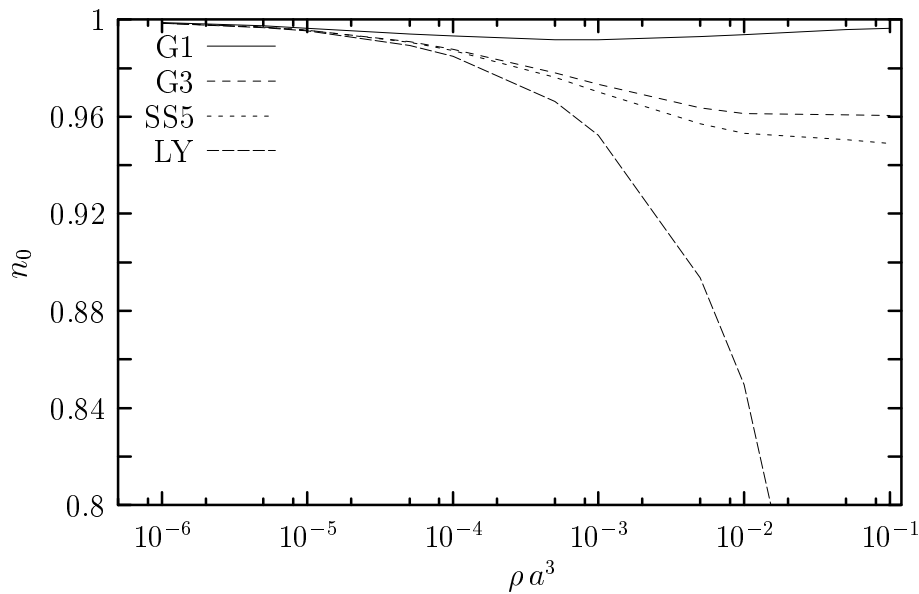}
\caption{\label{f7}
Condensate fraction of a weakly interacting Bose
gas as a function of the density for the potentials G1, G3 and SS5.
The low density expression of Eq.~(\ref{nqbog}) 
is also reported for comparison.
}
\end{figure*}

The condensate fraction $n_0$ is displayed in Fig. \ref{f7} for the
potentials G1, G3 and SS5 as a function of $\rho a^3$, 
and compared with the low--density 
expansion \cite{bogoliubov47,beliaev58,huang}

\begin{eqnarray}
n_0 = 1-\frac{8}{3}\sqrt{\frac{\rho a^3}{\pi}} \ .
\label{nqbog}
\end{eqnarray}

One can see that the condensate fraction of the IPC model starts
flattening at $\rho a^3 \approx 10^{-2}$, as for the case of $E/\rho$.
However, it remains below one up to the highest density value
considered. 

\begin{figure*}
\includegraphics{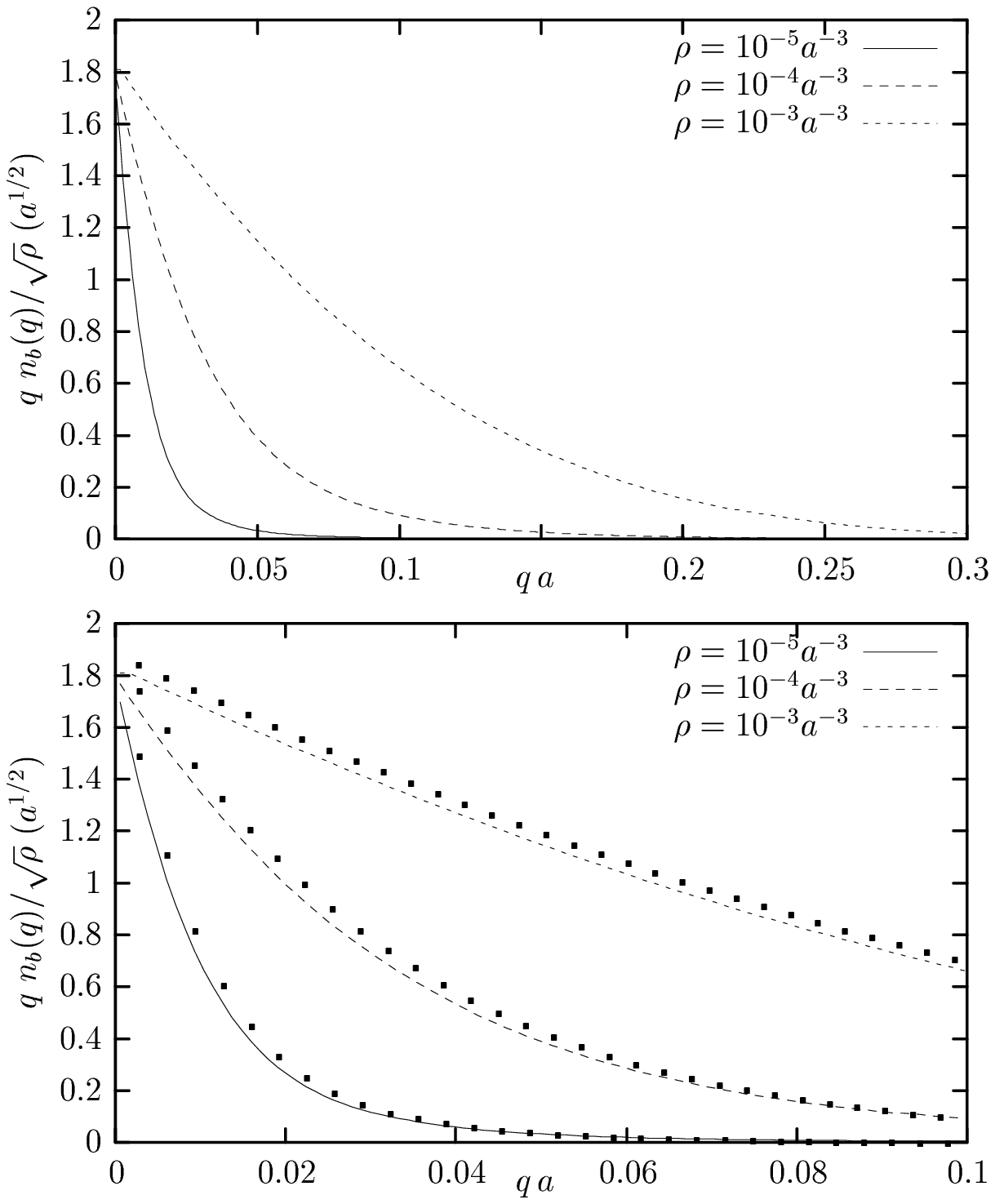}
\caption{\label{f8}
Momentum distribution  of the IPC  model of a Bose
gas interacting with the potential SS10 at several densities.
The momentum distribution $n_d(q)$
is multiplied by $q$ to make clear the $q^{-1}$ long wavelength behavior
and divided by $\sqrt{\rho}$ to fit the scale.
The full circles in the lower panel 
indicate the Bogoliubov results, $\sqrt{E_{un}}/2$.
}
\end{figure*}

The depletion momentum distribution $n_d(k)$ for the potential SS10 at 
three density values is displayed in Fig. \ref{f8}. The optimized 
distribution has interaction induced $k \neq 0$ (pair) occupation
even at $T=0$, reminiscent of the Fermi distribution.

Following the Fermi analogue it is useful to formally define Bose 
wavevector $q_b$ and bose energy, $\epsilon_d$ and temperature $T_b$ 
through

\begin{equation}
n_d(q_b)\equiv \frac{1}{2} \ , ~~~~~
\epsilon_d \equiv
\frac
{\hbar^2 q_b^2}
{2 \, m}
\equiv 
k_B T_b
\end{equation}

The arrows in Fig. \ref{f8} denote the bose wavevectors for the densities
considered. For the typical values of the scattering length and mean
interparticle spacing, 50 \AA ~ and 2000 \AA ~ respectively, typical
of BEC \cite{leggett01}, we find that 
$\epsilon_p \approx 1.12 10^{-11}$ eV
and  $T_b \approx 0.134 \mu$ K.

To better show the $q^{-1}$ behavior in the long 
wavelength limit, the figure plots 
$q\, n_d(q)$ rather than $n_d(q)$. Similar results have been obtained
for the other four potentials.
The extrapolated values at $q \rightarrow 0$
are compared with the theoretical sum rule value 
$d_{-1}^{\text{SR}}= \sqrt{n_0^2 {\cal K}/4}$ where the
compressibility ${\cal K}$ is given by

\begin{equation}
{\cal K} =\frac{d}{d\rho}\left(\rho^2\frac{dE}{d\rho}\right)
         = mc^2 \ ,
\label{sound}
\end{equation}

\begin{figure*}
\includegraphics{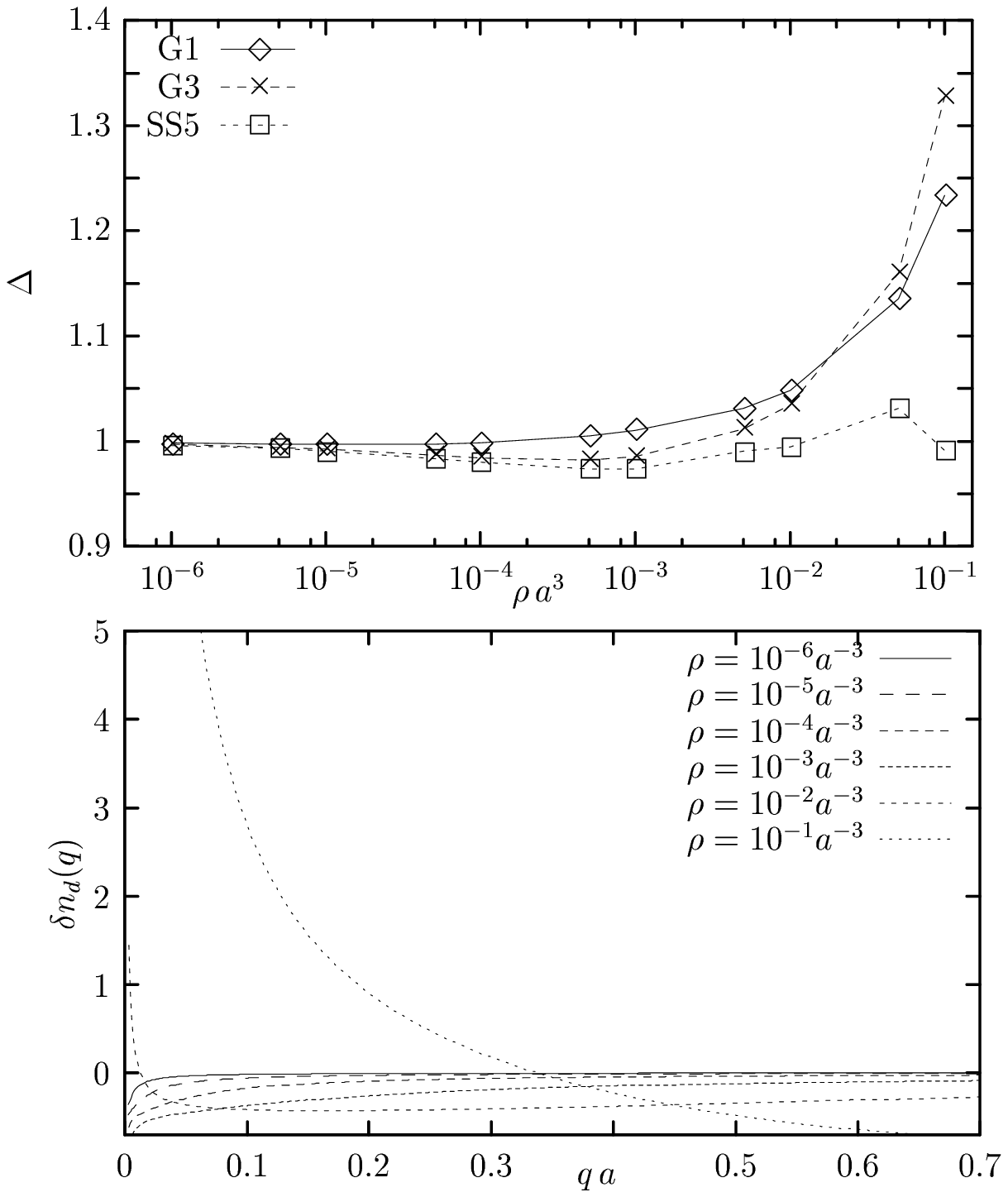}
\caption{\label{f9}
Compressibility sum rule. 
Upper plot: The quantity $\Delta$ of Eq.~(\ref{delta}) is
plotted as a function of the density for the potentials G1, G3 and SS5.
Lower plot: Difference function of the momentum distribution with respect
to that given for the exact sum rule at low $q \, a$ at different
densites for the $SS5$ potential.
}
\end{figure*}

We plot in Fig. \ref{f9} the quantity

\begin{eqnarray}
\Delta = \frac{d_{-1}}{d_{-1}^{\text{SR}}}
\label{delta}
\end{eqnarray} 

as a function of $\rho a^3$ for the potentials G1, G3 and SS5.
The compressibility sum rule is satisfied when $\Delta =1$.
For the sake of clarity we also plot in the lower panel 
the following difference  function

\begin{equation}
\delta n_d(q)=n_d(q)-\frac{d_{-1}^{\text{SR}}}{q}.
\end{equation}

We have found that over a range of densities 
$\rho \, a^3 \lesssim 10^{-3}$ 
the compressibility sum rule is satisfied within three 
significant figures.

The coefficient $d_{-1}$ can also be calculated directly, by using the 
following expression, which results from the terms in $q^{-1}$ of the 
Euler equation (\ref{euler_new}).

\begin{eqnarray}
d_{-1}^2 = \frac{n_0\tilde{v}_0+\int \frac{d\vec{q}}{(2\pi)^3\rho}
\tilde{v}(q)\widetilde{P}(q)}{8\left(1+\frac{1}{6}
\int\frac{d\vec{q}}{(2\pi)^3\rho}
\left(n_d(q)-\widetilde{P}(q)\right) \nabla^2_q\tilde{v}(q)\right)}\ .
\end{eqnarray}

The coefficient $d_{-1}$ obtained from the above equation and 
$\lim_{q\rightarrow 0} q n_d(q)$ coincide within three digits
for all the cases considered in the present calculation.

\begin{figure*}
\includegraphics{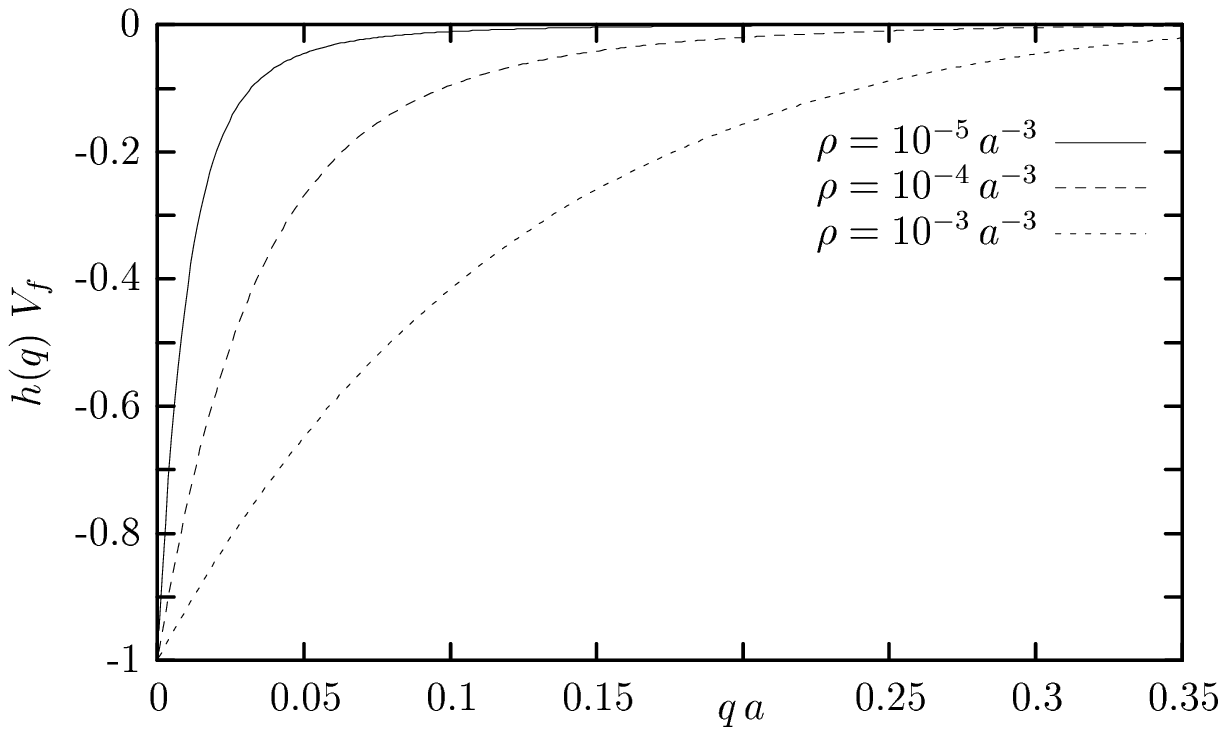}
\caption{\label{f10}
Pair correlation function of the IPC model. 
The table displays the Fourier transforms of the pair
correlation functions $\tilde{h}(q)$ vertex corrected by $V_f$, 
obtained for the case of the potential SS10.
}
\end{figure*}

We have also computed the chemical potential defined by 

\begin{eqnarray}
\mu &=& E+\rho\frac{dE}{d\rho} \ .
\label{chemical}
\end{eqnarray}

The results obtained for the sound velocity and for the chemical potential
are given in Table \ref{table5} for the potential G1, which has the ratio
$a/a_{\text{Born}}$ closer to $1$. The
corresponding estimates resulting from the low density expansion of 
Eq.~(\ref{energyLY}), and given by 

\begin{eqnarray}
(mc^2)_{LY} &=& \left(\frac{\hbar^2}{2ma^2}\right) 
8\pi\rho a^3\left( 1+ 16\sqrt{\frac{\rho a^3}{\pi}}\right)\ , \\
(\mu)_{LY} &=& \left(\frac{\hbar^2}{2ma^2}\right) 
8\pi\rho a^3\left( 1+ \frac{32}{3}
\sqrt{\frac{\rho a^3}{\pi}}\right)\ , 
\label{soundLY}
\end{eqnarray}

are also reported in the table.

\begin{table}
\caption[]{\label{table5}
Sound velocity $c$ and chemical potential $\mu$ at various density values
for the potential G1. The low--density results given in eqs. (\ref{soundLY})
are also reported for comparison. The density is givne in units of
$a^{-3}$ and the compressibility, $mc^2$, and chemical potential, $\mu$,
in units of $\hbar^2/(2ma^2)$
}
\begin{ruledtabular}
\begin{tabular}{lcccc}
$\rho$ & $(mc^2/\rho)_{LY}$ & $(mc^2/\rho)_{IPC}$ & $(\mu/\rho)_{LY}$ 
& $(\mu/\rho)_{IPC}$  \\
\hline
 $10^{-6}$ & $25.36$& $25.35$ & $25.28$ & $25.28$ \\
 $10^{-5}$ & $25.85$& $25.67$ & $25.61$ & $25.52$ \\
 $10^{-4}$ & $27.40$& $26.11$ & $26.65$ & $25.92$ \\
 $10^{-3}$ & $32.31$& $26.39$ & $29.92$ & $26.28$ \\
 $10^{-2}$ & $47.82$& $26.45$ & $40.26$ & $26.42$ \\
 $10^{-1}$ & $96.88$& $26.46$ & $72.96$ & $26.45$ \\
\end{tabular}
\end{ruledtabular}
\end{table}
The pair correlation function $h(r)$ multiplied by the vertex correction 
$V_f$ can be calculated from the momentum distribution $n_d(q)$, 
by using Eq.~(\ref{background}). Its
Fourier transform is plotted versus $q$ in Fig. \ref{f10} at
three different values of $\rho a^3$ for the potential SS10.

\begin{figure*}
\includegraphics{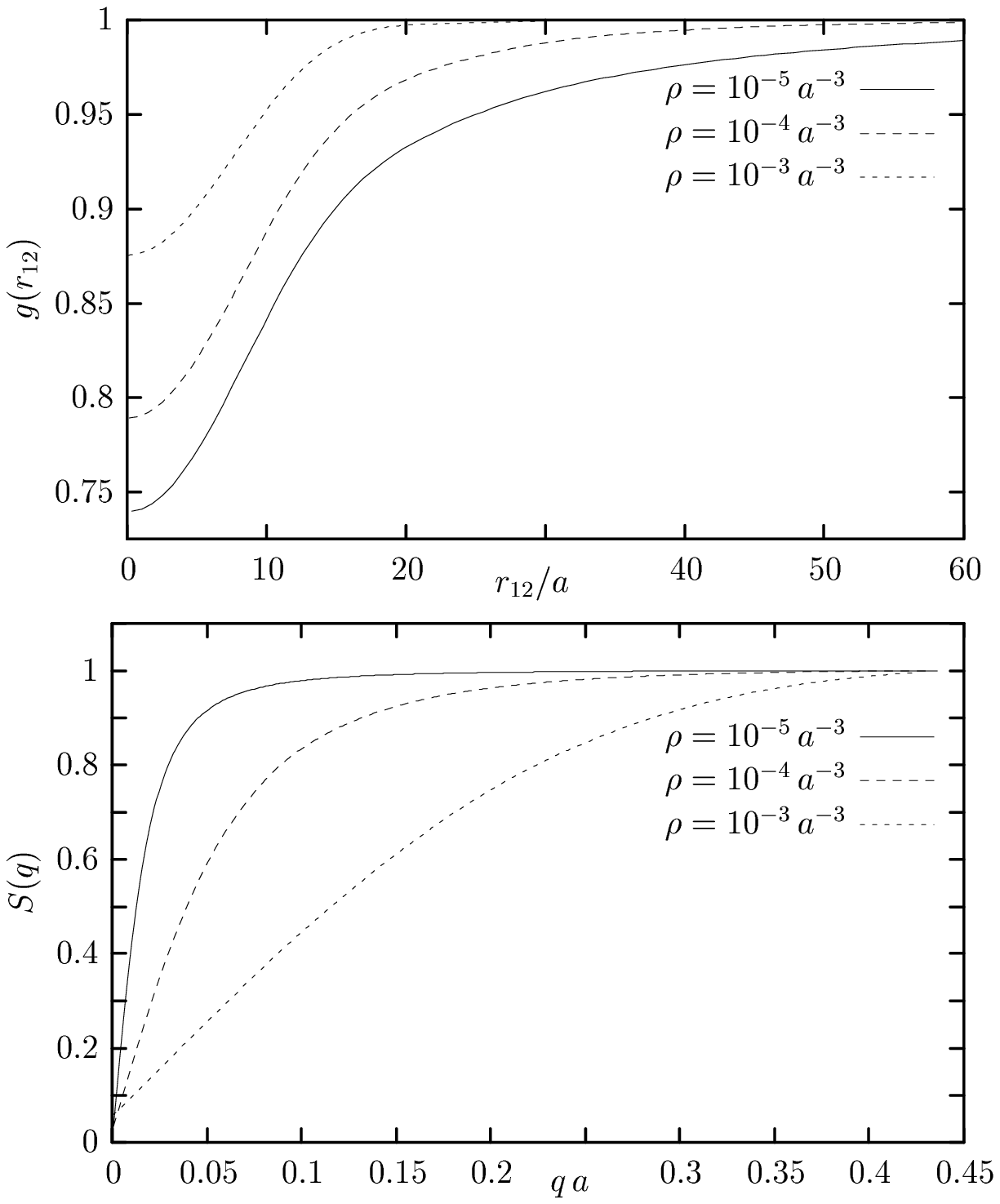}
\caption{\label{f11}
Upper plot:
Two--body distribution function $g(r_{12})$ of the IPC  model of a Bose
gas interacting with the potential SS10.
Lower plot: the same for the static structure factor $S(q)$.
}
\end{figure*}

\begin{figure*}
\includegraphics{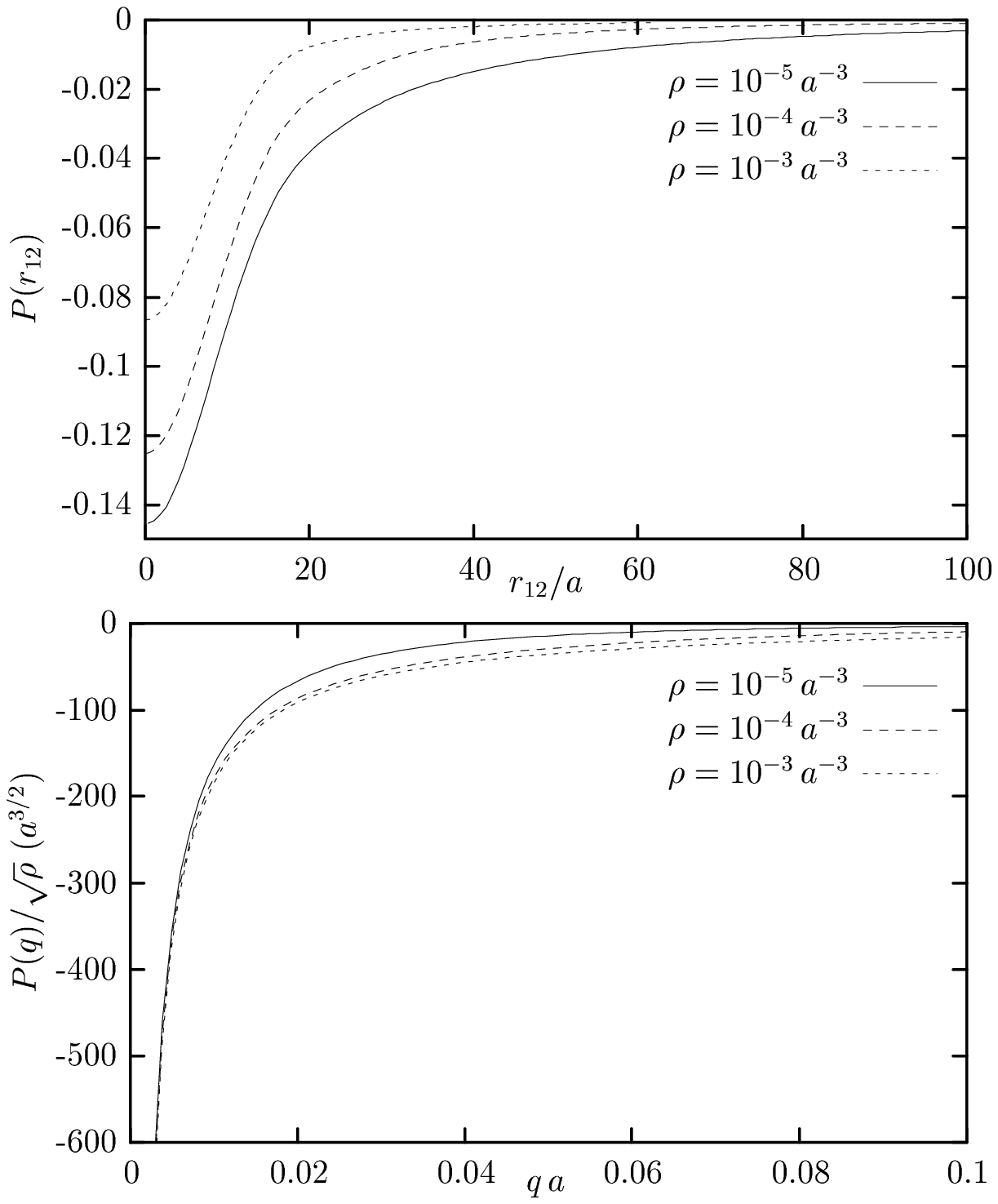}
\caption{\label{f12}
Upper plot:
Pairing function $P(r_{12})$ of the IPC  model of a Bose
gas interacting with the potential SS10.
Lower plot: the same for the pairing function in momentum, 
$\widetilde{P}(q)$, space divided by $\sqrt{\rho}$ to fit the scale.
}
\end{figure*}

Finally, the two--body distribution function 
$g(r_{12})$ and the pairing function 
$P(r_{12})$ are displayed in Figs. \ref{f11} and \ref{f12} respectively, 
for the case of the potential SS10. Similar plots can be shown for 
the other potentials. 
The long wavelength limits of the static structure function,
$S(0^+)$, obtained in correspondence with 
the $g(r_{12})$ reported in Fig \ref{f11} are small, 
$0.0102$, $0.0270$ and $0.0551$ for
$\rho a^3 = 10^{-5}, 10^{-4}, 10^{-3}$ respectively. 
The values extrapolated directly 
from $S(q)$ and those
computed from Eq.~(\ref{lwsq}) coincide within three digits for all 
the cases considered.

The pairing function $P(r_{12})$, as the one--body density
matrix $\rho(r_{11'})$, is long--ranged. 
Its Fourier transform. $\widetilde{P}(q)$, has therefore a singular 
long wavelength
behavior, given by $-n_0 mc/(2q)$, opposite to that of the momentum
distribution $n_d(q)$. Therefore, the function 
$F_1(r) = P(r)+\rho(r)$ is a short--range function. This function is 
the key ingredient of the following important  
ODLRO property of the semi--diagonal two--body density
matrix \cite{stringari95,ristig89} 

\begin{eqnarray}
\lim_{r_1'\rightarrow\infty}\rho(1,2;1',2) &=& 
n_0(1+F_1(r_{12})) \ .
   \end{eqnarray}

One can easily verify that $\widetilde{F}_1(0) = -1/2$, 
as required by the Stringari sum rule on the two--body 
density matrix \cite{stringari95}. Such a  behavior of
$\widetilde{F}_1(q)$ has important consequences on the overlap between 
the particle and density excited states \cite{shenoy02}. Moreover the sum
rule relationg $\mu$ to $F_1$ \cite{stringari95} is satisfied numerically
to three significant figures \cite{shenoy02} impliying gapless exitation.

\section{Conclusions and perspectives}

In this paper we have used the variational theory to develop a 
number--conserving model for boson pairing based upon the 
Bogoliubov theory. This model consider an 
independent pair correlated wave function as trial function, which   
is characterized by a $100\%$ condensation of $(\vec{k},-\vec{k})$ pairs.

The cluster expansions for the two--body distribution function 
$g(r_{12})$ and
the one-- and two--body density matrices from the IPC trial function 
have been derived, and a scheme of the HNC type has been developed to 
sum up the cluster terms to all order in a fully
closed form. This novel HNC method maps the Bose system into a Fermi system,
such that the fermions have as many flavours 
as the number of particles, and then it makes use of the 
Renormalized FHNC theory to compute the resulting cluster terms. 

As a consequence, the energy expectation value 
can be obtained with no approximation
for any given pair correlation $h(r)$, resulting in a true energy
upperbound. Similarly, the normalization sum rules on $g(r_{12})$, 
$\rho(r_{11'})$  and $\rho_2(1,2;1',2')$, as well as the kinetic energy
sum rule, are exactly satisfied.  
This cannot be achieved in Jastrow theory, where truncations of the cluster
series are intrinsically made.

The Euler equation to get the optimal pair correlation $h(r)$, obtained in
accordance with the Ritz variational principle, has been derived. Since 
the total energy per particle can be written in terms of the momentum
distribution only, we could derive the integro--differential 
equation having the momentum
distribution as variable, instead of $h(r)$. 
We found that the Euler equation in $n_d(q)$ is not 
uniquely determined, and we selected the one consistent with a momentum 
distribution which has the Gavoret--Nozi\'{e}res singularity in the long
wavelength limit. We proved that such an Euler equation can be 
numerically solved by means of a rather simple iterative self--consistent
procedure, and that the resulting $n_d(q)$ has indeed the required
long wavelength behavior. 

Calculations have been performed for five different pseudopotentials, 
three of the gaussian form, and two of the soft--sphere type, spanning
a region  of interaction range going from $\sim 0.1 r_0$ to $\sim 10 r_0$ 
and of strength $a/r_0$ from $\sim 0.02$ to $\sim 0.8$.

We found that IPC results for the energy per particle and the
condensate fraction are in reasonably good agreement 
with the available DMC calculations,
particularly for those potentials for which the scattering length $a$ is
closer to its value $a_{\text{Born}}$ in Born approximation. 
In the limit of $a/a_{\text{Born}} \rightarrow 1$ the low density 
expansion, Bogoliubov theory, DMC calculations and IPC model all give 
very close results. However, even for $a/a_{\text{Born}} \sim 0.7$, 
IPC provides a quite realistic  representation  of the corresponding 
Bose system. 

We were also able to compute other 
interesting quantities, like the sound velocity, the chemical potential
and the pairing function. The last quantity is directly related to the
off--diagonal long range order in the gas, and to its superfluid
properties at zero temperature. Therefore the IPC model appears to be
an excellent candidate to study boson pairing in a quantitative way.

We also found that the range of validity of the low--density expansion
is limited to $\rho a^3 \sim 10^{-4}$. Accordingly, for values of 
$\rho a^3$ of the order or larger than $10^{-3}$, 
the results for the various quantities of 
interest do dipend significantly on the shape of the potential and
not only on the scattering length $a$.

The proposed theory can be easily extended to 
treat finite Bose systems, typical of
trapped atomic vapours produced to study BEC and macroscopic coherent
phenomena. Work in this direction is in progress.

The generalization to Fermi systems or to Fermi--Bose mixtures
is also possible, although less straightforward. 
In this case, contrary to the Bose case, one has to deal with the
elementary diagrams. In the Fermi case, the generic 
point of a cluster diagram can be reached by up to four
bonds (instead of two), two of the $h$--type and 
two of the statistical type. This feature 
still garantees a closed set of FHNC integral equations, but with
kernels given by four--body functions.

A third, challenging perspective is the study of the excitation spectrum
and of the superfluid properties. Such studies draw on the IPC physical
picture of a number-conserving ground state with both a condensate and
a coherent zero-momentum paired depletion;
and on the HNC treatment developed in this paper,
which allows for exact evaluation of the expectation value of a generic
n--body operator. 

\begin{acknowledgments}

The authors are indebted to A.~Fabrocini and J.~Navarro for fruitful
discussions. T.M.N. would like to thank the 
International School for Advanced Studies in 
Trieste for partial support during the completion of this work. The
support from the grant MIUR-9902623127 (1999) 
is gratefully acknowledged by S.F. and A.S.
\end{acknowledgments}

\appendix*

\section{}

In this Appendix we derive the Euler equation resulting from the 
Ritz variational principle expressed by Eq.~(\ref{eulerq}).

Let us compute the functional derivatives of the vertex corrections $S_2$ 
and $V_f$. The vertex correction $V_b$ can be expressed in terms of $V_f$.

\begin{eqnarray}
\frac{\delta S_2}{\delta\tilde{h}(q)} & = & 
C \frac{\delta V_f}{\delta\tilde{h}(q)} + \frac{q^2}{\pi^2\rho} F(q)\ , 
\nonumber \\
\frac{\delta V_f}{\delta\tilde{h}(q)} & = & -V_f^2 \left(
\frac{\delta S_2}{\delta\tilde{h}(q)} + 
\frac{2\tilde{h}_0}{(1-\tilde{h}_0 V_f)^3}
\frac{\delta V_f}{\delta\tilde{h}(q)}\right) \ ,
\label{varver}
   \end{eqnarray}

with

\begin{eqnarray}
C & = & \int \frac{d\vec{k}}{(2\pi)^3\rho} 
\frac{\tilde{h}^2(k)(1+\tilde{h}^2(k) V_f^2)}
   {(1-\tilde{h}^2(k) V_f^2)^2}\ ,  \nonumber \\
F(q) & = & \frac{\tilde{h}(q) V_f} {(1-\tilde{h}^2(k) V_f^2)^2}\ .
\label{Ccorr}
   \end{eqnarray}

The solution of the above linear system is given by

\begin{eqnarray}
\frac{\delta S_2}{\delta\tilde{h}(q)} & = & 
\frac{q^2}{\pi^2\rho} \left( \frac{a_1}{a_1+C} \right) F(q)\ ,
\nonumber \\
\frac{\delta V_f}{\delta\tilde{h}(q)} & = &
-\frac{q^2}{\pi^2\rho} \left( \frac{1}{a_1+C} \right) F(q)\ ,
\nonumber \\
a_1 & = & \frac{2\tilde{h}_0}{(1-\tilde{h}_0 V_f)^3} +
\frac{1}{V_f^2}\ .
\label{varversol}
   \end{eqnarray}

The functional variation of the expectation value of the kinetic energy
results to be

\begin{eqnarray}
& & \frac{\delta\langle T \rangle}
{\delta\tilde{h}(q)}  =  \left(\frac{q^2}{\pi^2\rho}\right)
\frac{\hbar^2}{2m}V_f q^2 F(q)  \\
& + & \frac{\delta V_f}{\delta\tilde{h}(q)} 
\left( 2\frac{\langle T\rangle}{V_f} +
\frac{\hbar^2}{m} \int \frac{d\vec{k}}{(2\pi)^3\rho}
\frac{k^2\tilde{h}^4(k) V_f^3)} 
   {(1-\tilde{h}^2(k) V_f^2)^2} \right) \ , \nonumber
    \end{eqnarray}

which leads to 

\begin{eqnarray}
\frac{\delta\langle T\rangle}{\delta\tilde{h}(q)} & = & 
\left(\frac{q^2 V_f}{\pi^2\rho}\right)
\frac{\hbar^2}{2m} \left( q^2 F(q) - 2 F(q) T_1 \right) \ , \nonumber \\
T_1 & = & \frac{1}{a_1+C} \int \frac{d\vec{k}}{(2\pi)^3\rho}
\frac{k^2\tilde{h}^2(k)} {(1-\tilde{h}^2(k) V_f^2)^2} \ .
\label{varT}
    \end{eqnarray}

The functional variation of the potential energy expectation value 
$\langle V\rangle$ is given by

\begin{eqnarray}
\frac{\delta\langle V\rangle}{\delta\tilde{h}(q)}  =  
\left(\frac{q^2 V_f}{\pi^2\rho}\right)
F(q)\left(
\frac{\tilde{V}_b(q)}{\tilde{h}(q)}+\tilde{V}_a(q)-V_1 \right) \ ,
\label{varV}
\end{eqnarray}

where $\tilde{V}_a(q)$ and $\tilde{V}_b(q)$ are defined as 

\begin{eqnarray}
 & &\tilde{V}_a(q) =\left( \tilde{v} | \frac{\tilde{h}^2 V_f^2}
{1-\tilde{h}^2 V_f^2}\right)_q \ , \nonumber \\ 
 & & \tilde{V}_b(q)=\frac{1}{2} \left(
\frac{(1+\tilde{h}(q) V_f)^2}{(1-\tilde{h}_0 V_f)^2} \tilde{v}(q)
\right. \nonumber \\
& & \left. +
(1+\tilde{h}^2(q) V_f^2)
 \left( 
\tilde{v} | \frac{\tilde{h}}{1-\tilde{h}^2 V_f^2}\right)_q\right)\ ,
\label{termVq}
\end{eqnarray}

and the constant term $V_1$ by

\begin{eqnarray*}
V_1 & = & \frac{1}{a_1+C} \left(\frac{1+\tilde{h}_0 V_f}
{(1-\tilde{h}_0 V_f)^3}
\int \frac{d\vec{k}}{(2\pi)^3\rho}
\frac{\tilde{h}(k)\tilde{v}(k)} {1-\tilde{h}(k) V_f} \right.
\\
&+& \frac{1}{(1-\tilde{h}_0 V_f)^2}
\int \frac{d\vec{k}}{(2\pi)^3\rho}
\frac{\tilde{h}(k)\tilde{v}(k)}{(1-\tilde{h}(k) V_f)^2} 
\\
&+& 2 \int \frac{d\vec{k}}{(2\pi)^3\rho}\tilde{v}(k)
\left(\frac{\tilde{h}^2}{(1-\tilde{h}^2 V_f^2)^2}|
\frac{\tilde{h}^2 V_f^2}{1-\tilde{h}^2 V_f^2} \right)_k 
\\
&+& \left. \int \frac{d\vec{k}}{(2\pi)^3\rho} \tilde{v}(k)
\left(\frac{\tilde{h} (1+\tilde{h}^2 V_f^2)}
{(1-\tilde{h}^2 V_f^2)^2}|
\frac{\tilde{h}}{1-\tilde{h}^2 V_f^2} \right)_k\right) \ . 
\end{eqnarray*}

We can now write down explicitly the Euler equation 
resulting from  Eq.~(\ref{eulerq})

\begin{eqnarray}
F(q) \left( \frac{\hbar^2}{2m}(q^2-2T_1)+
\frac{\tilde{V}_b(q)}{\tilde{h}(q)}+\tilde{V}_a(q)-V_1 \right) = 0 \ .
\label{EULER}
   \end{eqnarray}

Since $F(q)$ cannot be equal to zero, then the Euler equation becomes

\begin{eqnarray}
\frac{\hbar^2}{2m}q^2\tilde{h}(q)+
\left(\tilde{V}_a(q)-\lambda\right)\tilde{h}(q)+\tilde{V}_b(q) = 0\ ,
\label{EULER1}
\end{eqnarray}

with

\begin{eqnarray}
\lambda = \frac{\hbar^2}{m} T_1 + V_1 \ .
\label{lambda}
   \end{eqnarray}

The quantities $\tilde{V}_a(q)$, $\tilde{V}_b(q)$ and $\lambda$ all 
depend on $h(r)$ in a highly non
linear way. Therefore eqs.(\ref{EULER1}) and (\ref{lambda}) 
need to be solved by means of a self consistent iterative procedure.
The equation (\ref{EULER1}) is equivalent to Eq.~(\ref{euler_new}),
as one can verify by expressing $\tilde{h}(q)$ in terms of $n_d(q)$.


\begin{thebibliography}{99}

\bibitem{BEC}
M. Anderson, \emph{et al.}
Science \textbf{269}, 198 (1995); 
C. C. Bradley, \emph{et al.}
Phys. Rev. Lett. \textbf{75}, 1687 (1995); 
K. B. Davis, \emph{et al.}
\emph{ibid.} \textbf{75}, 3969 (1995); 
M. O. Mewes,  \emph{et al.}
\emph{ibid.}  \textbf{77}, 416 (1996); 

\bibitem{dalfovo99}
F. Dalfovo,  \emph{et al.}
Rev. Mod. Phys. \textbf{71}, 463 (1999).

\bibitem{ketterle99}
W. Ketterle, D. S. Durfee, and D. M. Stamper--Kurn, in 
\emph{Bose--Einstein Condensation in Atomic Gases}, 
Proceedings of the  International School of Physics "Enrico Fermi",
Course CXL, edited by M. Inguscio, S. Stringari, and C. Wieman,
(IOS Press, Amsterdam, 1999). 

\bibitem{leggett01}
A. J. Leggett, 
Rev. Mod. Phys. \textbf{73}, 307 (2001).

\bibitem{bogoliubov47}
N. N. Bogoliubov, 
J. Phys. (USSR) \textbf{11}, 23 (1947); 
reprinted in D. Pines, \emph{The Many Body Problem} 
(Benjamin, New York, 1962).

\bibitem{huang} 
K. Huang, 
\emph{Statistical Mechanics}, 
(John Wiley \& Sons, New York, 1987).

\bibitem{feenberg}
E. Feenberg, 
\emph{Theory of Quantum Fluids}, 
(Academic Press, New York, 1969).

\bibitem{girardeau59}
M. D. Girardeau, and R. Arnowit, 
Phys. Rev. \textbf{113}, 755 (1959).

\bibitem{valatin58}
J. G. Valatin and D. Butler, 
Nuovo Cimento \textbf{10}, 37 (1958).

\bibitem{evans69}
W. A. B. Evans, and Y. Imry, 
Nuovo Cimento B \textbf{63}, 155 (1969).

\bibitem{shenoy77}
S. R. Shenoy, and A. C.Biswas, 
J. of Low Temp. Phys. \textbf{28}, 191 (1977).

\bibitem{bohm67}
D. Bohm, B. Salt, 
Rev. Mod. Phys. \textbf{39}, 894 (1967).

\bibitem{pines59}
N. M. Hugenholtz, and D. Pines, 
Phys. Rev. \textbf{116}, 489 (1959).

\bibitem{henshaw61}
D. G. Henshaw and A. D. B. Woods, 
Phys. Rev. \textbf{121}, 1266 (1961).

\bibitem{nozieres99}
P. Nozi\'eres, D. Pines, 
\emph{The Theory of Quantum Liquids} 
(Perseus Books Publishing, L. L. C., 1999)

\bibitem{gardiner97}
C. W. Gardiner, 
Phys. Rev. A \textbf{56}, 1414 (1997).

\bibitem{legget69}
A. J. Legget in 
\emph{Modern trends in the Theroy of Condensed Matter},
edited by A. Pekolski and R. Przystowa,
(Springer Verlag, Berlin 1980);
J. Phys. (Paris) Colloq. \textbf{41}, 456 (1969).

\bibitem{randeria95}
M. Randeria in 
\emph{Bose--Einstein Condensation}, edited by 
E. Griffin, D. W. Snoke and S. Stringari, 
(Cambridge University Press, Cambridge, 1995).

\bibitem{blatt64}
J. M. Blatt, 
\emph{Theory of superconductivity},
(Academic, New York, 1964).

\bibitem{nozieres64}
J. Gavoret, and P. Nozi\'eres, 
Ann. of Phys. \textbf{28} 349 (1964) 349;
L. Reatto and G. V. Chester, Phys. Rev. \textbf{155}, 88 (1967).

\bibitem{stringari95}
S. Stringari, in 
\emph{Bose--Einstein Condensation}, edited by 
E. Griffin, D. W. Snoke and S. Stringari, 
(Cambridge University Press, Cambridge 1995).

\bibitem{fantoni74}
S. Fantoni, and S. Rosati, 
Nuovo Cimento A \textbf{20} 179 (1974).

\bibitem{fantoni98}
S. Fantoni, and A. Fabrocini, in
\emph{Microscopic quantum-many body theories and their applications},
Lecture Notes in Physics, Vol. \textbf{510}, 
edited by J. Navarro and  A. Polls,
(Springer-Verlag, Belin, 1998).

\bibitem{navarro02}
J. Navarro, in 
\emph{Advances in Quantum Many--Body Theory}, 
edited by A. Fabrocini, S. Fantoni and E. Krotscheck,
World Scientific, in press.

\bibitem{navarro02A}
J. Navarro, private communication.

\bibitem{aldrich76}
C. H. Aldrich III, and D. Pines, 
J. of Low Temp. Phys. \textbf{25}, 677 (1976).

\bibitem{moroni95}
S. Moroni, S. Fantoni, G. Senatore, 
Phys. Rev. B \textbf{52}, 13547 (1995).

\bibitem{fantoni78}
S. Fantoni, 
Nuovo Cimento A \textbf{44}, 191 (1978).

\bibitem{pandhaLOCV}
V. R. Pandharipande, 
Nucl. Phys. A \textbf{174}, 641 (1971) 641; 
\emph{ibid.}  \textbf{178}, 123 (1971); 
V. R. Pandharipande and H. A. Bethe,
Phys. Rev. C \textbf{7}, 1312 (1973).

\bibitem{ristig79}
M. L. Ristig, 
Nucl. Phys. A \textbf{317}, 163 (1979).

\bibitem{ristig80}
M. L. Ristig, and P. M. Lam, 
J. of Low Temp. Phys. \textbf{40}, 571 (1980).

\bibitem{fantoni81}
S. Fantoni, 
Nucl. Phys. A \textbf{363} 381 (1981).

\bibitem{baym69}
G. Baym, in 
\emph{Mathematical Methods Solid State and Superfluid Theory},
edited by R. C. Clark, G. H. Derrick, 
(Oliver \& Boyd, Edimburgh, 1969).

\bibitem{lee57}
T. D. Lee and C. N. Yang, 
Phys. Rev. \textbf{105}, 1119 (1957);
T. D. Lee, K. Huang and C. N. Yang, 
Phys. Rev. \textbf{106}, 1135 (1957).

\bibitem{cornish00}
S. L. Cornish, \emph{et al.},
Phys. Rev. Lett. \textbf{85}, 1795 (2000).

\bibitem{cowell01}
S. Cowell, \emph{et al.},
e-print  cond--mat/0106628.

\bibitem{giorgini99}
S. Giorgini, J. Boronat, and J. Casulleras, 
Phys. Rev. A \textbf{60}, 5129 (1999).

\bibitem{joachain79}
C. J. Joachain, 
\emph{Quantum collision theory},
(North Holland, Amsterdam, 1979).

\bibitem{lieb98}
E. H. Lieb, and J. Yngvason, 
Phys. Rev. Lett. \textbf{80} 2504 (1998).
E. H. Lieb, R. Seiringer, and J. Yngvason, 
Phys. Rev A \textbf{61}, 043602 (2000).

\bibitem{fabrocini99}
A. Fabrocini and A. Polls, 
Phys. Rev. A \textbf{60} 2319 (1999);
{\bf 64},  063610 (2001)

\bibitem{wu59}
T. T. Wu, 
Phys. Rev. \textbf{115} 1930 (1959).

\bibitem{beliaev58}
S. T. Beliaev, JETP \textbf{34}, 433 (1958).

\bibitem{ristig89}
M. L. Ristig and J. W. Clark, 
Phys. Rev. B \textbf{40}, 4355 (1989).

\bibitem{shenoy02}
S. R. Shenoy, A. Sarsa, T. M. Nguyen and S. Fantoni, in preparation.

\end{thebibliography}
\end{document}